\documentclass[a4paper,11pt]{article}
\usepackage{jcappub} 
\usepackage{lineno}
\usepackage{amsmath}
\usepackage{bm}
\usepackage{hyperref}
\usepackage[flushleft]{threeparttable}
\usepackage{booktabs} 
\usepackage{float} 
\usepackage[numbers]{natbib}
\usepackage{cancel}
\usepackage{ulem}
\usepackage{subcaption}
\usepackage{graphicx}
 \usepackage{xcolor}
\title{\boldmath  From ANN to BNN: Inferring reionization parameters using uncertainty-aware emulators of \\21-cm summaries}

\author[a]{Yashrajsinh Mahida,}
\author[a]{Sanjay Kumar Yadav,}
\author[a]{Suman Majumdar,}
\author[a]{Leon Noble,}
\author[a,b,c]{Chandra Shekhar Murmu,}
\author[d,e]{Saswata Dasgupta,}
\author[f]{Sohini Dutta,}
\author[g,h]{Himanshu Tiwari,}
\author[i]{Abinash Kumar Shaw}

\affiliation[a]{Department of Astronomy, Astrophysics \& Space Engineering,\\ Indian Institute of Technology Indore,\\ Indore 453552, India}

\affiliation[b]{Astrophysics Research Centre of the Open University (ARCO),\\ The Open University of Israel,\\ 1 University Road, POBox808, Ra’anana 4353701, Israel}

\affiliation[c]{Department of Natural Sciences,\\ The Open University of Israel,\\ 1 University Road, POBox808, Ra’anana 4353701, Israel}

\affiliation[d]{Institute of Astronomy,\\ University of Cambridge, Cambridge, UK}

\affiliation[e]{Kavli Institute for Cosmology,\\ University of Cambridge, Cambridge, UK}

\affiliation[f]{
Jodrell Bank Centre for Astrophysics, University of Manchester, Manchester M13 9PL, UK}

\affiliation[g]{International Centre for Radio Astronomy Research (ICRAR),\\ Curtin University,\\ Bentley, Perth, WA, Australia}

\affiliation[h]{Commonwealth Scientific and Industrial Research Organisation (CSIRO),\\ Space \& Astronomy,\\ Bentley, WA, Australia}

\affiliation[i]{Department of Computer Science, University of Nevada Las Vegas, 4505 S. Maryland Pkwy., Las Vegas, NV 89154, USA}

\emailAdd{yhmahida@gmail.com}

\abstract{Inferring astrophysical parameters from radio interferometric observations of the redshifted 21-cm signal from the Epoch of Reionization (EoR) is a challenging, yet crucial. A Bayesian inference pipeline for reionization parameter estimation, forward models the signal statistic, usually the power spectrum, and compares it to the observed statistic. However, the 21-cm signal coming from EoR is expected to be highly non-Gaussian in nature; therefore, we need to use higher-order statistics, e.g., bispectrum. Moreover, the forward modeling of the signal and its statistics for a varying set of model parameters requires rerunning the numerical simulations a large number of times, which is computationally very expensive and time-consuming. To overcome this challenge, many artificial neural network (ANN) based emulators have been introduced, which produce the 21-cm signal summary statistics in a fraction of the time, given input astrophysical parameters. However, ANN emulators have a fundamental drawback: they can only produce point-value predictions; thus, they fail to capture the uncertainty associated with their own predictions. Therefore, when such emulators are used in the Bayesian inference pipeline, they cannot naturally propagate their prediction uncertainties to the estimated model parameters. To address this problem, we have developed Bayesian neural network (BNN) emulators for the 21-cm signal statistics, which provide the posterior distribution of the predicted signal statistics, including their own prediction uncertainty. We use these BNN emulators in our Bayesian inference pipeline to infer the EoR parameters through 21-cm summaries of the mock observation of 21-cm signal with telescopic noise for $1000$ hr of SKA-LOW observation. We show that BNN emulators are able to capture the prediction uncertainty for the 21-cm power spectrum and bispectrum, and using these emulators in the inference pipeline provides better and tighter constraints on them. To check the robustness of the emulators, we systematically reduced the training dataset and showed that, for smaller training datasets, BNN outperforms the ANN emulators. We also show that using the bispectrum as a summary statistic gives better constraints on EoR parameters than the power spectrum.
}


\begin{document}
\maketitle
\flushbottom

\section{Introduction}

The goal of modern cosmology is to understand the evolutionary history of the Universe as it has gone through different epochs of evolution. The phase of the Universe, when the first luminous sources started to form, is known as Cosmic Dawn (CD). The subsequent phase, in which the radiation from these luminous sources begins to ionize the neutral hydrogen (HI) of the intergalactic medium (IGM), is known as the Epoch of Reionization (EoR). These epochs are still not very well understood. Our current understanding of CD-EoR prominently comes from different indirect observations such as the Cosmic Microwave Background (CMB) \cite{Komatsu_2011, Planck_2016, Planck_2018}, the absorption spectra of high redshift quasars \cite{Becker_2001, Fan_2003, Barnett_2017}, and the luminosity function and clustering properties of Lyman-$\alpha$ emitters \cite{Ouchi_2010, Zheng_2017, Taylor_2021}, which cannot provide solid answers to essential questions about CD-EoR such as exact duration of these epochs, formation and properties of the ionizing sources, the morphology of ionized regions and its evolution. We require direct observations of CD-EoR to understand these epochs better. 

Multiple efforts are going on for the direct observation of the EoR using the 21-cm emission line as a signal. As most of the IGM is made up of neutral hydrogen, the redshifted 21-cm signal from the hyperfine spin-flip transition of neutral hydrogen works as a tracer of the ionized region during EoR \cite{Furlanetto_2006, Pritchard_2012}. There are many ongoing radio interferometers, such as uGMRT \cite{Paciga_2013}, HERA \cite{HERA_2022a}, LOFAR \cite{Mertens_2025}, MWA \cite{MWA_2019}, and the upcoming SKAO \cite{Koopmans_2015}, that are capable of detecting the 21-cm signal from EoR. Even though we are waiting for the direct detection of this signal, we already have upper limits on the spherically averaged 21-cm power spectrum from these interferometers \cite{Paciga_2013, Gehlot_2019, MWA_2019, Merterns_2020, HERA_2022a, HERA_2022b, Acharya_2024, Ceccotti_2025, Mertens_2025}. Recently, \cite{Gill_2025} has provided the upper limits on the 21-cm bispectrum from MWA observations. 

The redshifted 21-cm signal coming from the EoR is known to be highly non-Gaussian \cite{Bharadwaj_2005, Iliev_2006, Mellema_2006}. This non-Gaussianity arises due to the underlying non-linear matter density distribution traced by the ionizing sources, and the evolving ionized regions. The power spectrum, which gives the variance of the signal at different $k$-modes, is an optimal statistic to describe any Gaussian field completely. Since the redshifted 21-cm signal is highly non-Gaussian, the power spectrum alone cannot capture all the information available in the signal. To extract the non-Gaussian information present in this signal, we need higher-order statistics such as the bispectrum \cite{Majumdar_2018, Majumdar_2020, Kamran_2021a, Kamran_2021b, Kamran_2022, Mondal_2021, Shimabukuro_2016, Hutter_2020, Watkinson_2019, Ma_2021, Raste_2024, Noble_2024, Watkinson_2022, Tiwari_2022}. Bispectrum, the Fourier conjugate of the three-point correlation function, has more information about the field than the power spectrum since it can capture the correlations present between the different scales. Therefore, it can provide more information about the astrophysical processes that drive the evolution of this non-Gaussianity. This extra information can help to put tighter constraints on the astrophysical parameters of the EoR models \cite{Shimabukuro_2017b, Tiwari_2022}. In \cite{Tiwari_2022}, it is demonstrated that considering the bispectrum for all unique triangle configurations as a summary statistic can provide tighter constraints on the EoR model parameters compared to the power spectrum.

One of the primary goals of the EoR observations through different radio interferometric arrays is to estimate the reionization model parameters from the observed redshifted 21-cm signal statistics through a Bayesian inference framework\cite{Mondal_2020, Ghara_2020, Ghara_2021, Greig_2021, HERA_2022b, Ghara_2025}. Such a parameter estimation pipeline functions by having many random walkers scout the unknown multi-dimensional parameter space by estimating the target observable statistics at every step, which can typically be within the range of $10^4 - 10^6$ steps. Generally, one uses radiative transfer or seminumerical simulations to model these observable statistics; however, these simulations can be computationally very expensive and time-consuming to use in such a Bayesian inference pipeline. One way to overcome this bottleneck is to use neural network-based emulators for the target statistics. Once these emulators are trained on a sufficiently large and optimally sampled simulated signal database to predict the target signal statistic, they can predict the same for an unknown set of parameters in a few seconds. Thus, such emulators can be an ideal candidate to use in the inference pipelines where we need fast prediction of the observables iteratively. Recently, there have been several works on developing neural networks to emulate different EoR 21-cm signal statistics \cite{Schimt_2018, Jennigs_2019, Tiwari_2022, Sikder_2024, Breitman_2024}. Additionally, there have been efforts towards developing neural networks to directly predict reionization model parameters from signal statistics or signal maps \cite{Shimabukuro_2017, Gillet_2019, Chodhury_2020, Prelogovic_2022, Choudhury_2022, Tripathi_2024, Choudhury_2025}. Most of the neural networks used in these works have a fundamental drawback, i.e., they can only produce point values of the target statistics. Further, Neural networks are prone to overfitting, and thus, it is very difficult to tune hyperparameters optimally for a given training dataset. Therefore, if a neural network is not trained robustly enough, it can make erroneous predictions without any warning flags. Since it can only predict point values, there is no direct way to quantify the uncertainty associated with its prediction. When such neural network architectures are used to make emulators used in the Bayesian inference pipeline, these prediction uncertainties are not naturally propagated to the estimation of the model parameters. One way to propagate these errors is using GPR as described in \cite{Kern_2017}. However, emulator errors propagated in this fashion are constant throughout the parameter space, which is not always true. Thus, it is very important to quantify these prediction uncertainties for every step in the parameter space.

To address these shortcomings of deterministic emulators, we have developed a Bayesian neural network (BNN) based emulator for EoR 21-cm observables. Bayesian neural networks use Bayesian inference to train the network for a given training dataset. The key difference between BNN and ANN is that a BNN learns the distribution of the weights and biases (i.e., network parameters) instead of point values, as is the case for an ANN. Thus, when we draw a prediction from a BNN, it gives us the distribution of the predicted quantity corresponding to the distribution of its tuned network parameters. Therefore, we can directly quantify the uncertainties associated with the network predictions. Further, we can propagate these prediction uncertainties through our Bayesian inference pipeline to get a more robust estimation of the reionization parameters. There have been a couple of works related to the use of BNN for the estimation of the EoR parameters \cite{Hortúa_2020, Meriot_2025}. Both of these works mainly focus on using the BNN for simulation-based inference (SBI) to directly estimate the EoR parameters. In this article, we train a BNN (via a database of simulated EoR 21-cm signal) to emulate the signal statistics and then use these emulators in an MCMC Bayesian inference pipeline to estimate the EoR parameters. \cite{Meriot_2025} has compared an ANN emulator-based MCMC pipeline with SBI to estimate the parameters. They found that the former systematically performs better than the latter (including BNN). We have developed a BNN-emulator-based parameter estimation pipeline and shown that the BNN-based emulator can provide better and more robust constraints compared to an ANN-based emulator in an MCMC inference pipeline because of the quantification of the prediction uncertainties. For this work, we have developed a BNN-based emulator for both the power spectrum and the bispectrum of the 21-cm signal. We compare the performance of the BNN emulator with the corresponding ANN emulator in constraining the EoR parameters. We also systematically reduce the size of the training dataset and compare the performance of both types of emulators for the reduced training datasets to understand the minimum amount of training data required to get robust estimations. As mentioned earlier, the redshifted 21-cm signal from the EoR is highly non-Gaussian; thus, the use of a higher-order statistic, e.g., bispectrum, can provide better constraints on the reionization parameters compared to the power spectrum (\cite{Tiwari_2022} has demonstrated it with an ANN emulator). Therefore, we compare the performance of the EoR 21-cm power spectrum and bispectrum in constraining the EoR model parameters when both of them are emulated by two robustly trained BNN emulators.

This article is organized as follows: In section \ref{Simulation}, we describe the details of the simulation used to generate the 21-cm signal maps used as the training dataset. Section \ref{Statistic estimation} describes how we estimate the signal statistics from the reionization maps. In section \ref{BS_ANNvsBNN}, the drawbacks of the ANN-based emulators and how BNN-based emulators help to overcome those drawbacks are discussed. In section \ref{Emulators}, we have described ANN and BNN emulator architectures and their performance for both power spectrum and bispectrum. In section \ref{Results}, we compare the results of parameter estimations provided by ANN and BNN emulators. Finally, in section~\ref{Summary}, we summarize and discuss our findings.

\section{Reionization 21-cm Signal Simulation}{\label{Simulation}}
We use a semi-numerical simulation based on excursion set formalism \cite{Furlanetto_2004} to generate the redshifted 21-cm brightness temperature field of the EoR. This process involves three main steps. First, we generate the dark matter (DM) density field using a dark matter-only particle-mesh (PM) $N$-body code \cite{Bharadwaj_Srikant_2004, Mondal_2015} at redshift $z=7$. We simulated our DM density field on $3072^3$ grids with the spatial grid resolution of $0.07$ cMpc (comoving Mpc), which makes our simulation volume $[215.04~\rm {cMpc}]^3$ and DM particle mass resolution of $1.09 \times 10^8 \rm{M_\odot}$. We use the Friends-of-Friends (FoF)~\cite{Davies_1985, Mondal_2015} algorithm to identify the collapsed halos within the distribution of DM particles using a fixed linking length of $0.2$ times the mean interparticle distance. The criterion to identify the collapsed halo is that it should have at least $10$ DM particles, which results in a minimum halo mass of $1.09 \times 10^9 \rm{M_\odot}$. For the final step, we apply an excursion set-based algorithm\cite{Furlanetto_2004} to the DM field and the halo catalog to generate the ionization field using the semi-numerical code \texttt{ReionYuga}\footnote{\url{https://github.com/rajeshmondal18/ReionYuga}} \cite{TRC_2009, Majumdar_2014, Mondal_2017}. This code assumes that the hydrogen density perfectly follows the underlying DM density and that all the DM halos host the sources that produce the ionizing photons. There are three free parameters in this simulation framework: 1) $\bm{M_{\rm h,\, min}}$:  This parameter sets the lower cutoff for the mass of the halo that participates in the reionization process. Only halos with a mass higher than $M_{(\rm h, min)}$ have the sources that produce ionizing photons, 2) $\bm{N_{\rm ion}}$: We consider that the number of ionizing photons produced by the sources are proportional to the mass of the host DM halo and this dimensionless proportionality constant is $N_{\rm ion}$. The $N_{\rm ion}$ is a parameter that essentially quantifies the efficiency of the ionizing sources. 3) $\bm{R_{\rm mfp}}$: This parameter represents the mean free path of ionizing photons in the IGM. For a detailed discussion of the reionization parameters, interested readers may refer to \cite{TRC_2009, Majumdar_2014, Mondal_2017}. All the length scales mentioned in this article are in comoving units, so hereafter we drop the cMpc notation in favor of Mpc.
\par
We generate our training data set by varying these parameters in the following range - $M_{\rm h,m in} ( \times 10^8 \rm{M_\odot}) \in [10, 800]$, $N_{\rm ion} \in [10, 200]$, and $R_{\rm mfp}$ (Mpc) $\in [1.12, 40.32]$. We use a uniform grid sampling (please refer to \cite{Tiwari_2022} to know more about the effect of sampling) in this 3D parameter space corresponding to the given parameter ranges, which allows us to sample a maximum of 7200 parameter sets, for which we simulate the EoR 21-cm brightness temperature maps at redshift $z=7$. Note that when comparing with a realistic observation via a radio interferometric array, one should ideally simulate the signal for the whole range of target frequency channels (or redshifts) to include the lightcone effect \cite{Datta_2012, Datta_2014, Mondal_2021} and for a volume representing the observing volume constituted by the field-of-view of the array multiplied with the comoving depth determined by the frequency or redshift range. This will require one to simulate the signal for a large number of coeval cubes at different redshifts within the observing redshift range and then constitute the signal lightcone by stitching the slices from coeval cubes. Performing this task for $7200$ parameter sets will require a significant amount of computing time and storage space for those coeval and lightcone cubes. Thus, for the purpose of the development of these emulators, we focus on a specific redshift and simulate one coeval cube centred at that redshift while varying the EoR parameters for building the signal maps for the training and validation dataset. This ensures that we capture the variation of the astrophysical parameters within our training set while not going overboard with the computing overhead. In the future, we wish to upgrade our emulators by training them on signal lightcones with adequate computing resources.

\section{Estimating Signal Statistics}{\label{Statistic estimation}}

\subsection{Power Spectrum}

The power spectrum is one of the most popular Fourier statistics to describe any field. For the 21-cm signal, the power spectrum can be expressed as:

\begin{align}
    \langle \Delta_{b}(\bm{k}) \Delta_{b}(\bm{k})^* \rangle = V \; P(\bm{k}),
\end{align}
where $\Delta_{b}(\bm{k})$ is the Fourier transform of the brightness temperature field, and $V$ is the volume of the simulation box. This power spectrum gives the power of the field for a particular $k$-mode. We estimate the bin-averaged power spectrum from our 3D simulation box by considering a spherical bin of radii between $k$ and $k+dk$. Thus, power spectrum of the $i^{th}$ bin is expressed as:

\begin{align}
{\bar{P}}_{i}(k)=\frac{1}{V\,N_k} \sum_{\bm{k}~\epsilon~i}    \langle \Delta_{b}(k) \Delta_{b}(-k) \rangle,
\end{align}
where $\rm{N_{k}}$ is the total number of $k$ modes inside the $i^{th}$ bin. For this work, we have used the normalized dimensionless power spectrum expressed as,
\begin{align}
    \Delta^2 (k) = \Big( \frac{k^3}{2\pi^2} \Big)~\bar{P} (k)
\end{align}

\subsection{Bispectrum}
The bispectrum is the Fourier conjugate of the three-point correlation function. For the 21-cm brightness temperature field, the bispectrum $B({k_1}, {k_2}, {k_3})$ is defined as:

\begin{align}
    \langle \Delta_{b}({k_1}) \Delta_{b}({k_2}) \Delta_{b}({k_3}) \rangle = {V} \delta_{\bm{k_1} + \bm{k_2} + \bm{k_3},0}~{B}({k_1},{k_2},{k_3}),
\end{align}
where $\Delta_{b}(\bm{k})$ is the Fourier transform of the brightness temperature field, and $V$ is the volume of the simulation box. The Kronecker delta $ \delta_{\bm{k_1} + \bm{k_2} + \bm{k_3}, 0} $ equals to one when the condition $\bm{k_1} + \bm{k_2} + \bm{k_3} = 0$ is met, which ensures that only closed $k$-triangles contribute to the bispectrum; otherwise, the Kronecker delta is zero. 

We estimate the binned bispectrum,
\begin{align}
{\bar{B}}_{i}(\bm{k_1},\bm{k_2},\bm{k_3})=\frac{1}{V\,N_{\rm tri}} \sum_{[\bm{k_1}+\bm{k_2}+\bm{k_3}=0]~\epsilon~i}    \langle \Delta_{b}(\bm{k_1}) \Delta_{b}(\bm{k_2}) \Delta_{b}(\bm{k_3}) \rangle,
\end{align}
from the 3D simulation box following the algorithm presented in~\cite{Majumdar_2018, Majumdar_2020}.
Here, $\rm{N_{tri}}$ represents the total number of statistically independent closed $k$-triangles in the $i^{\text{th}}$ bin. The bispectrum depends on the shape and size of the triangles in Fourier space. We parametrize our bispectrum via two additional parameters that help to characterize the shape of the triangle in the Fourier space: 
\begin{align}
    n = \frac{k_2}{k_1}, \quad \text{and} \quad \cos~\theta = -\frac{\bm{k_1} \cdot \bm{k_2}}{k_1 k_2},
\end{align}
where $k_1$ and $k_2$ are the magnitudes of the vectors  $\bm{k_1}$ and  $\bm{k_2}$, respectively, and $\theta$ is the angle between them. To estimate the bispectrum of the field, we are identifying the triangles in the Fourier space, with side lengths being the magnitude of the three wave vectors. There can be multiple triangles with the same shapes, depending on the values of these vectors. To identify all the unique triangle configurations, we apply the following uniqueness conditions presented in~\cite{Bharadwaj_2020}
\begin{gather}
    k_1 \geq k_2 \geq k_3,\\
    0.5 \leq n \leq 1.0,\\
    0.5 \leq \cos{\theta} \leq 1.0~.
\end{gather}
Similar to the power spectrum, we use a normalized dimensionless bispectrum following~\cite{Majumdar_2020},
\begin{align}
    \Delta^3 (k_1,n,\cos{\theta}) = \bigg(\frac{k_1^6 n^3}{(2\pi^2)^2}\bigg)~{\bar{B}} (k_1,n,\cos{\theta}) .
\end{align}

\begin{figure}
    \centering
    \includegraphics[width=0.45\linewidth]{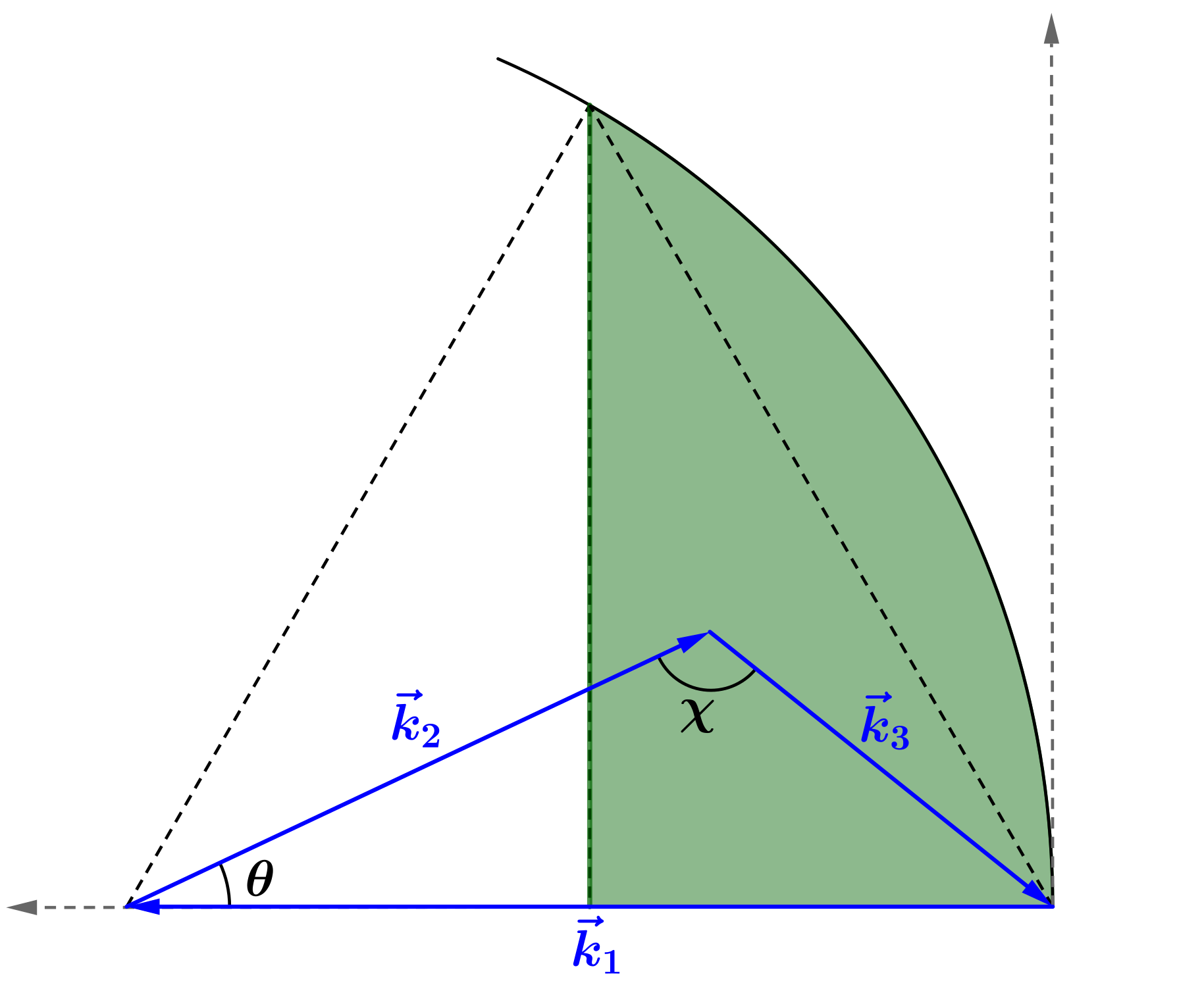}
    \includegraphics[width=0.45\linewidth]{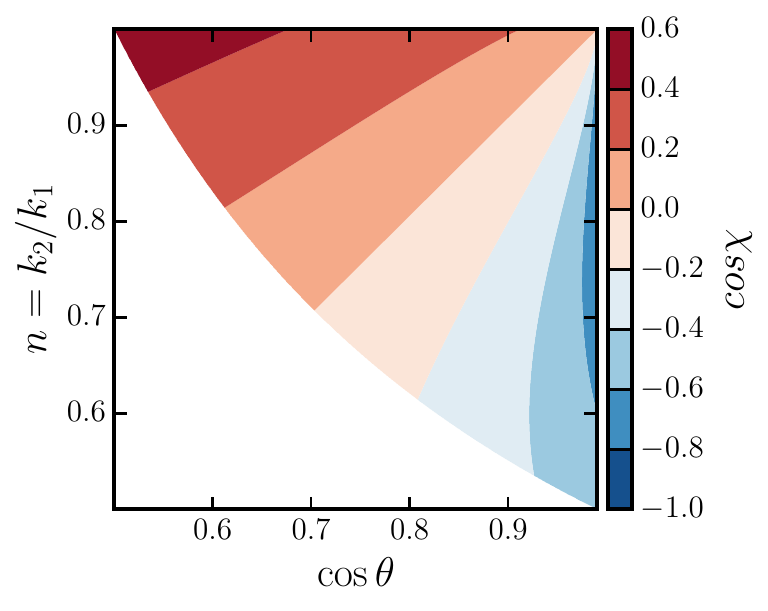}
    \caption{The left panel shows the allowed region (green shade) for ~$k_2$ spanning the triangles of all unique shapes for a fixed ~$k_1$. The right panel shows how the different unique shapes (denoted by $\cos \;\chi $) of the triangles are distributed over the $n - \cos \;\mathrm{\theta}$ plane.}
    \label{fig:bispec parametrization}
\end{figure}

The above parametrization of bispectrum ($ \Delta^3 (k_1,n,\cos{\theta})$) helps to find all possible unique shapes of closed triangles. Since the 21-cm signal is a non-Gaussian signal, unlike the second-order power spectra, bispectra try to quantify correlations between three different $\mathbf{k}$-scales in an intricate manner, which is non-trivial to explain. However, various shapes are expected to have different information content of the signal as well as varying impact of observational contamination. Recently, \cite{Gill_2025} has shown how the observational effects (e.g., thermal noise and residual foreground) affect the different triangle configurations of the bispectrum using MWA data and provided the upper limits on the bispectrum.

\section{Emulators: ANN vs BNN}{\label{BS_ANNvsBNN}}

\subsection{Artificial Neural Network (ANN)}

An artificial neural network (ANN) is the fundamental neural network architecture of Deep Learning. The fundamental building block of ANN architecture is called a ``Neuron'' or a ``Node''. Multiple neurons make a layer, and these fully interconnected layers make a neural network. ANN consists of three layers: Input, Hidden, and Output. Information is processed as it flows through layers of the network, with each neuron in each layer making decisions based on its input and contributing to the final output. The learning or training of NN consists of two processes, namely forward propagation and backward propagation.

In forward propagation, the training data is given to the network via the input layer and passes through the hidden layers to the output layer, generating an output. Every neuron in the network has a point value of weight and bias associated with it. Each neuron takes the input from all the neurons of the previous layer; then it takes the linear combination of all the values with the weight and bias of the corresponding neuron. The activation function takes this linear combination as input, and the output of the activation function goes to the next layer. The activation function introduces nonlinearity to the network. When the output layer receives the results from the preceding hidden layer, it predicts the output. This predicted output is compared to the true output using a \emph{loss function}. Our goal is to minimize the value of the loss function, i.e., the weight and bias of all neurons should be tuned in such a manner that the network is able to predict the output values with minimal loss of some desired quantity. To achieve this, based on the value of the loss function, one has to update the values of every weight and bias. This task is taken care of by the backward propagation.

In backward propagation, the loss (or difference) between predicted and actual output values of training data is calculated using some predefined loss function. The weight and bias of each neuron are updated during each iteration of the training, such that the loss tends to zero. The task of determining the updated values for each neuron is taken care of by the \emph{optimization algorithms}. One of the most common optimization algorithms is the gradient descent algorithm \cite{Ruder_2016}. The basic idea behind gradient descent is to update the parameters in the direction opposite to the gradient of the loss function with respect to the parameters. This is done because the gradient points toward the steepest increase of the function, so moving in the opposite direction should lead to a decrease in the function value. The gradients of the loss with respect to each weight and bias are computed using the chain rule of calculus, allowing the network to understand how each parameter contributes to the error. The algorithm then adjusts the weights and biases to reach the minimum of the loss function. After weights and biases have been changed, again, forward propagation happens, and the loss is calculated. According to this, the values of weight and biases are updated. This iterative process is called training the network. It continues until we get the values of weights and biases such that we achieve the minimum desired loss. For a detailed discussion of ANN, interested readers may refer to \cite{Goodfellow-et-al-2016}.

\subsubsection*{Drawbacks of ANN}
 ANN architecture and training process have many hyperparameters, and it is a challenging task to tune those hyperparameters to get the combination for which we can achieve the minimum loss. So, for ANNs, we find the set of hyperparameters for which we get minimum loss, and corresponding to that, we have minimum loss point estimate values for ANN model parameters (i.e., weights and biases of each neuron), which means we have a single point values of weight and bias for each neuron, tuned using the backpropagation algorithm. So, when we get a prediction from the ANN, it is also a single-value prediction. The problem with single-value prediction is that the predicted value depends on the quality of the data in your training dataset, and we do not know the uncertainty related to this prediction. The quantification of uncertainties related to deep learning models is very crucial \cite{Abdar_2021}. There are fundamental uncertainties related to deep learning models: Aleatoric and Epistemic uncertainty \cite{Kiureghian_2009, Hüllermeier_2021, Gawlikowski_2023}. Aleatoric uncertainty is inherent to the data itself. It comes from the underlying randomness or just the lack of information in the data. This type of uncertainty can not be reduced just by providing more training data to the network. On the other hand, Epistemic uncertainty is the uncertainty related to the model, which tries to explain or, in our case, tries to emulate the data itself. This comes from uncertainties in the neural network model architecture (mainly from choices of hyperparameters).

One aims to make an emulator for a given summary statistic and use it to do the Bayesian inference to constrain the reionization model parameters. Since ANNs give point value predictions, they do not quantify all the uncertainties associated with them, which we mentioned above. There is no way to know how good (or bad) the prediction is in unknown regions of parameter space, as the ANN does not provide the measure of uncertainty in its predictions, and it can confidently predict highly inaccurate values. Use of such an emulator for Bayesian inference may suffer from a lack of trust in the inferred parameters, as there are unknown uncertainties from emulators propagating in the inference pipeline over which we do not have any control. Therefore, we need models that have some measure of the uncertainty associated with their predictions.

\subsection{Bayesian Neural Network (BNN)}

To overcome the drawbacks of ANNs, we need to consider models capable of quantifying uncertainty related to their predictions. One way to do this is to incorporate Bayesian statistics into the deep learning model. The fundamental idea behind Bayesian statistics is to use a probabilistic approach to quantify the uncertainties in the inferred parameters. Therefore, we want to train neural networks through the Bayesian inference approach. In this approach, we consider network parameters, e.g., weights and biases, as random variables and try to infer them given some prior dataset \cite{blundell_15}. These types of neural networks are known as Bayesian neural networks (BNN) \cite{Arbel_2023, Charnock_2020, Jospin_2022, Olivier_2021}. BNNs have some apparent advantages over ANNs. They are trained through the Bayesian approach, so they naturally incorporate the probabilistic way to quantify uncertainties. They are less prone to overfitting and underfitting the data. Improvement of network architectures to develop a better emulator also requires understanding the type of uncertainties (i.e., aleatoric or epistemic) contributing to the prediction. This categorization of the uncertainties is possible for the Bayesian neural networks by decomposing their predictive variance into aleatoric and epistemic uncertainties. Interested readers are referred to \cite{Valdenegro-Toro_2022, Gawlikowski_2023} for more details about the decomposition of uncertainties in BNNs. Furthermore, due to their probabilistic nature of training, they can, in principle, be trained even for a very small training dataset without overfitting. For predictions, they provide a large epistemic uncertainty for the out-of-distribution (OOD) points instead of confidently giving the wrong outputs like ANNs. All of these make BNN trustworthy and reliable for making a robust emulator to perform parameter estimation.

In BNNs, the network weights and biases are considered random variables. Given some training dataset $\mathcal{D}: \{(x_1,y_1), (x_2,y_2),...,(x_n,y_n) \}$, where $(x,y)$ is a pair of input and output. Consider $w$ be the set of parameter values that belong to $\mathbb{R}^N$ space, $N$ = number of parameters. We set some prior $p(w)$ for the distribution of the parameters $w$. The posterior distribution of parameters following Bayes' rule will be:

\begin{equation}
    p(w|\mathcal{D}) = \frac{p(\mathcal{D}|w)~p(w)}{p(\mathcal{D})},
\end{equation}
where $p(\mathcal{D}|w)$ is the likelihood and $p(\mathcal{D})$ is the model evidence. For simplicity, we will assume a Gaussian prior for the parameters. So, after marginalizing the posterior $p(w|\mathcal{D})$, we will get the Gaussian distribution of each parameter in terms of their mean and variance.

The training problem of BNN is now reduced to a parameter inference problem. There are two main computational approaches to infer the posterior \cite{Jospin_2022, Arbel_2023}: 1) Sampling techniques like Markov Chain Monte Carlo (MCMC) and 2) Approximation methods like Variational Inference. Even though variational inference techniques are less expensive to perform than traditional MCMC methods, in this paper, we have used the MCMC method to train our BNNs simply because they are quite accurate compared to approximation methods. Our BNNs do not have very complex architectures, which makes them ideal candidates for training using MCMC without too much computational overhead. The main advantage of sampling methods is that when we increase the number of iterations, we asymptotically should converge to the true posterior. To train our BNN models, we use the No-U-Turn sampling (NUTS) technique \cite{Homan_2014}. The NUTS algorithm is an advanced version of the Hamiltonian Monte Carlo (HMC) sampling technique. The HMC sampling uses the Hamiltonian equations to explore the parameter space using a random variable and its auxiliary momentum vector \cite{2011hmcm.book..113N}. It requires a time step and a total number of steps to perform the numerical integration. The NUTS algorithm learns this step size during the burn-in and automates the whole calculation without deciding the step size at each iteration.

Once the training is done for a given dataset and the posterior distribution over the model weights, i.e., $p(w|\mathcal{D})$ is obtained, it can then generate the predictions for some new unknown input data point $x^*$ by marginalizing the likelihood with the posterior distribution. 

\begin{equation}
    p(y^*| x^*, \mathcal{D}) = \int p(y^*| x^*, w) p(w|\mathcal{D})~dw.
\end{equation}
As mentioned above, in this specific work, all parameters are in terms of the Gaussian mean and variance. Hence, we get a BNN prediction in terms of the mean and variance of the predicted quantity.

\section{Emulation of Signal Statistics}{\label{Emulators}}

To build the training and test sets for our emulators, we first estimate the target signal statistics, i.e., power spectrum and bispectrum, from the simulated 21-cm maps for $7200$ parameter sets as described in Sec \ref{Simulation}. Out of these $7200$ signal estimates, we set aside randomly selected $50$ samples as the test set to check the performance of our emulators, and out of the remaining data, $ 80\%$   is used for training, and $20\%$  is used to validate the corresponding network's performance. We have used the \texttt{KERAS} \footnote{\url{https://keras.io/}} \cite{chollet2015keras} library based on the \texttt{TensorFlow}\footnote{\url{https://www.tensorflow.org/}} \cite{tensorflow2015-whitepaper} package to develop the ANN-based emulators and \texttt{PYRO}\footnote{\url{https://pyro.ai/}} \cite{bingham2018pyro} probabilistic programming library based on the \texttt{PyTorch}\footnote{\url{https://pytorch.org/}} \cite{Paszke_2019_pytorch} package to develop BNN-based emulators. Initially, we train our emulators for the entire dataset and tune the network to enhance their prediction accuracy and relative errors. However, to test the robustness of these emulators, we gradually reduce the training data set volume and try to identify the minimum number of points in the parameter space that are required for getting an optimally trained emulator, both for the power spectrum and bispectrum. The training of both emulators is done using the NVIDIA A100-SXM4-40GB GPU. ANN emulators required $\sim 30$ mins and BNN emulators $\sim 2$ hrs to train on this GPU.
\subsection{Power spectrum emulation}

We have developed emulators to predict the 21-cm signal power spectrum for three input reionization parameters. The networks are trained to predict the power spectrum for $7\;k$-bins in the range $[0.11 - 2.75]\; \rm{Mpc^{-1}}$. This range is chosen in accordance with the range of $k$-modes that upcoming telescopes will be able to probe. Both ANN and BNN emulator architectures have $3$ nodes for three input parameters as the input layer and $7$ nodes corresponding to the power spectrum values for each $k$-bin in the output layer. The detailed architecture of the ANN-based emulator is provided in table \ref{tab:PS_ANN_BNN}. As one can see, the architecture of the ANN-based emulators has more layers and nodes compared to the BNN-based emulator, which means BNN-based emulators are able to capture more information out of data than the ANN-based emulators with a smaller architecture. We have used ELU activation for both ANN and BNN emulators. The ANN emulator is trained for $300$ epochs with a learning rate of $\sim 10^{-3}$. The BNN-based emulator uses Bayesian inference to train the emulator given a training dataset. So they have different sets of hyperparameters than the ANN emulators. For training the BNN, one needs to specify the total number of steps for the MCMC chain to run and the likelihood function. We use the multivariate Gaussian log-likelihood as the likelihood function and train our model for 1000 MCMC steps, with the initial 500 steps serving as warm-up steps. The number of steps is selected based on two main criteria: convergence and training time. More steps take a longer time to train the model. So, we found that this number of steps was enough to train the model properly. If we consider more steps, the quality of the model does not improve considerably compared to the increased training time. 

\begin{figure}
    \centering
    \includegraphics[width=0.8\linewidth]{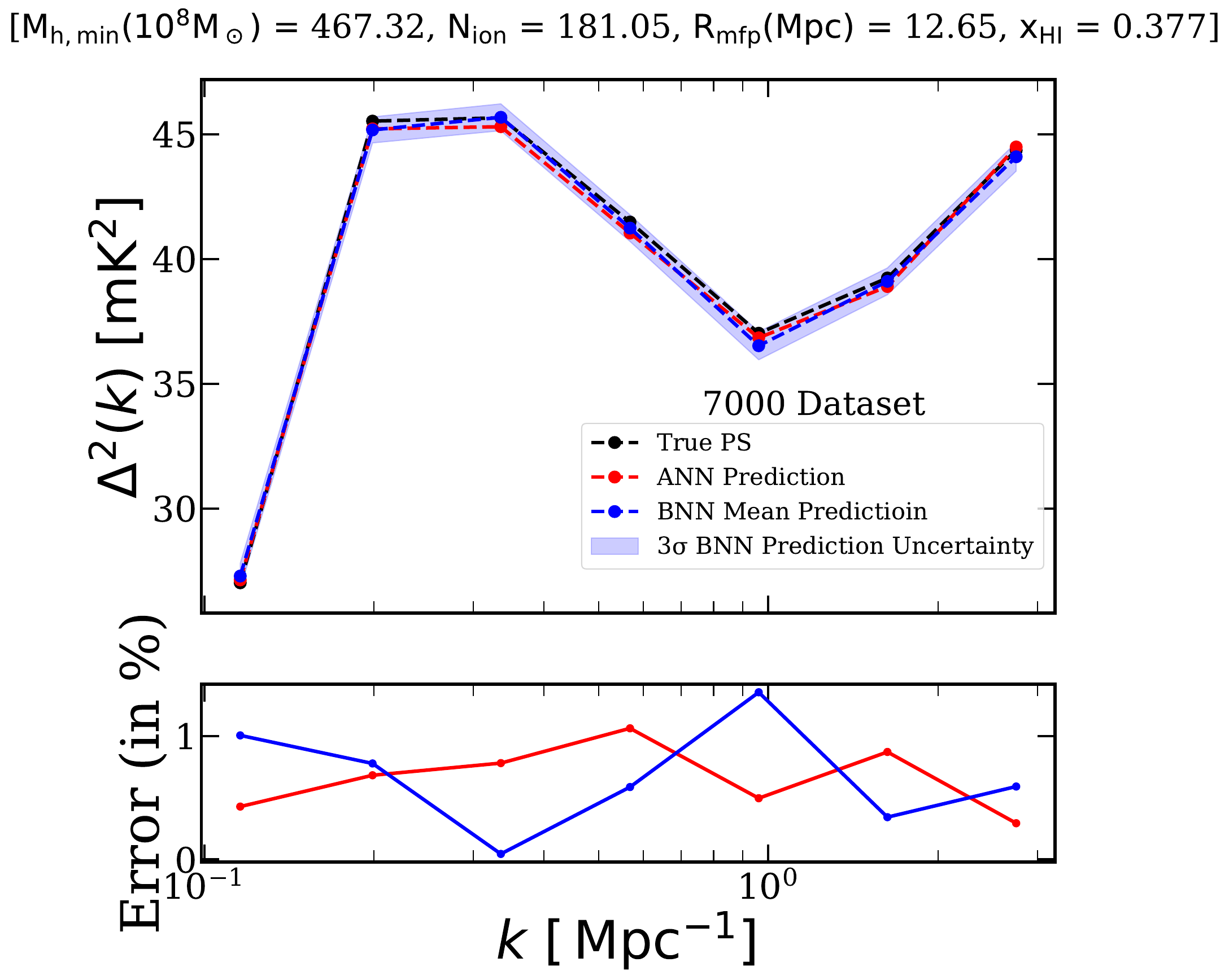}
    \includegraphics[width=0.50\linewidth]{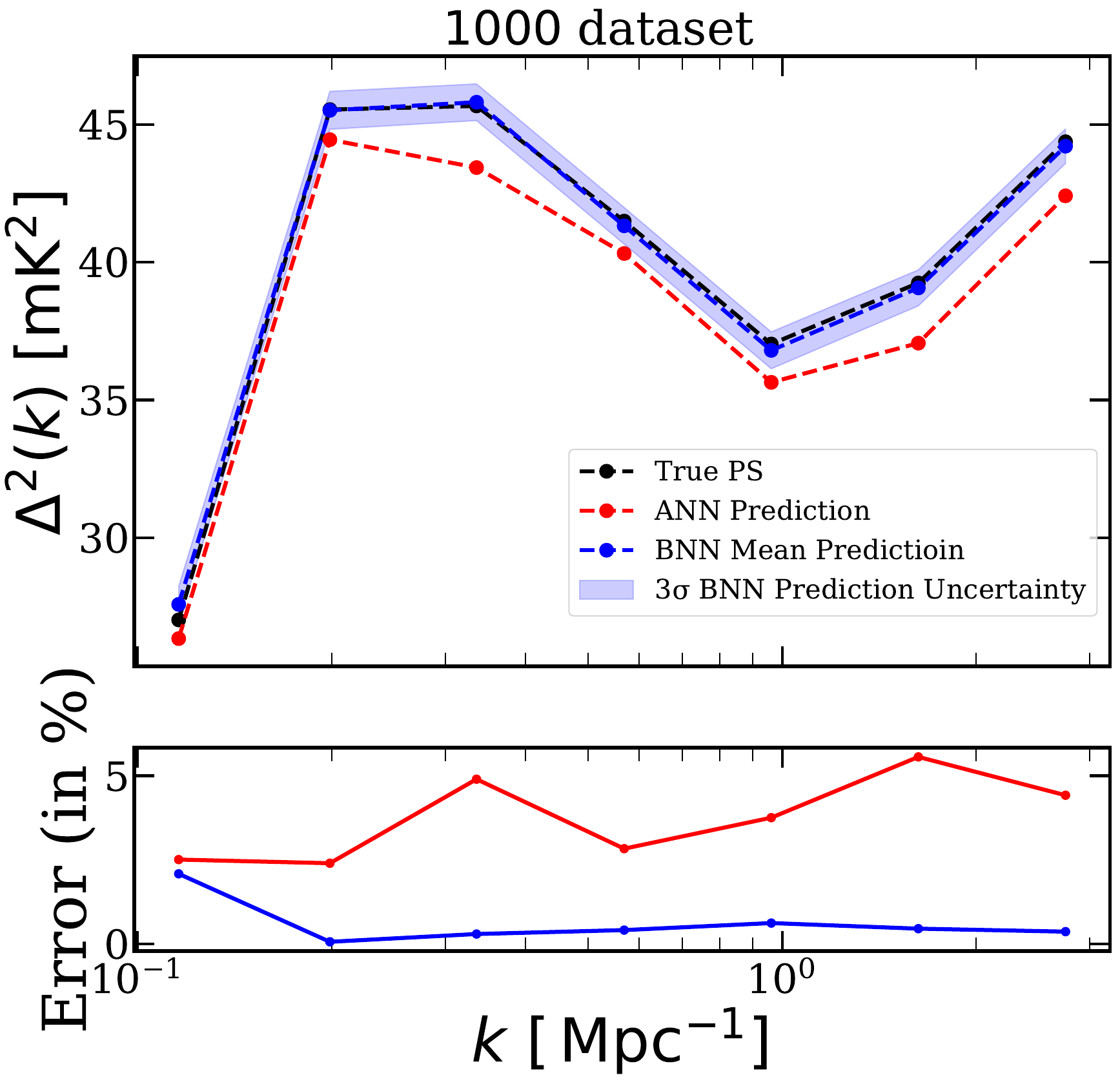}
    \includegraphics[width=0.475 \linewidth]{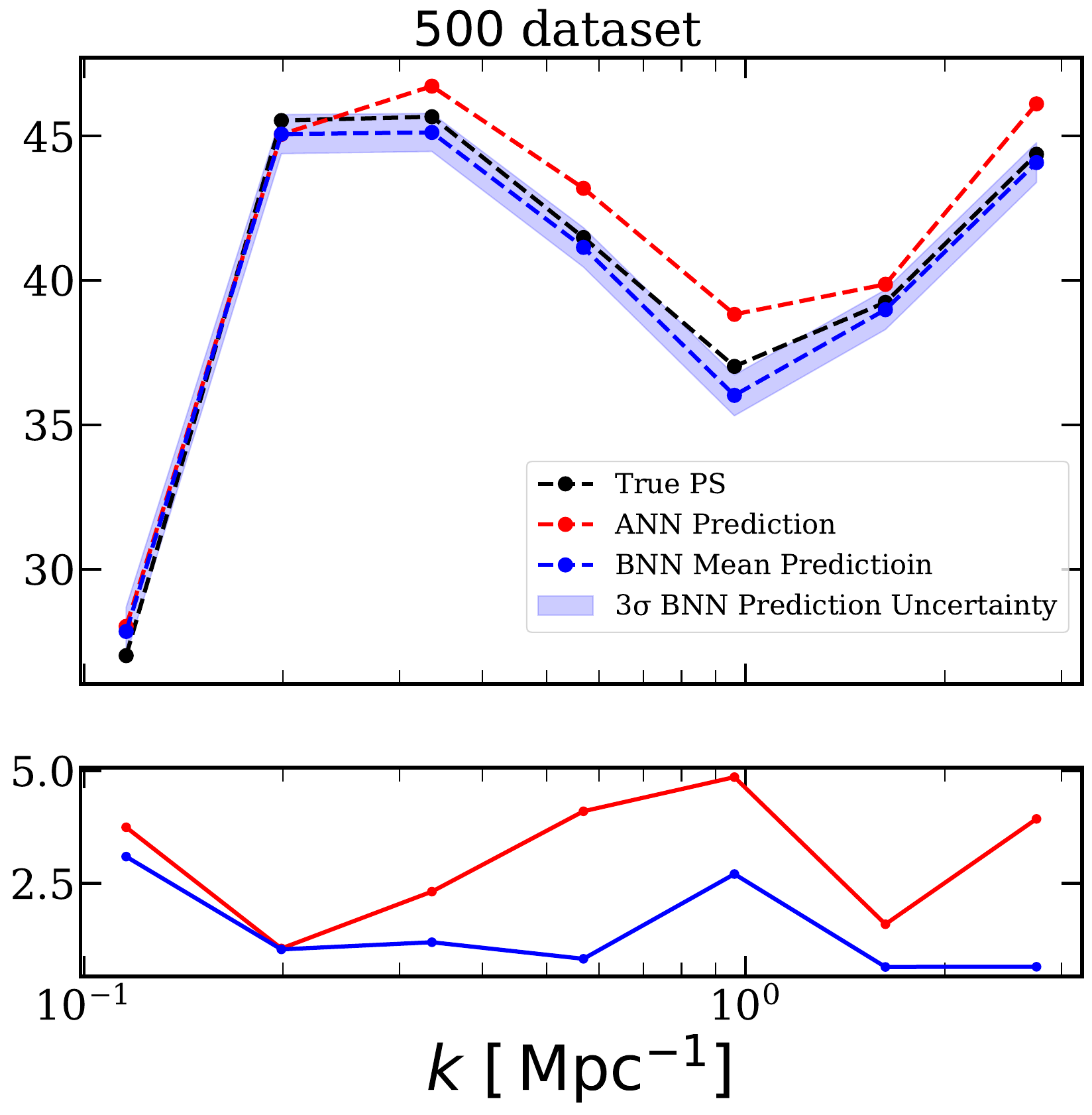}
    \caption{Plot shows the performance of ANN and BNN power spectrum emulators trained on different dataset sizes. The top plot shows the results for the complete dataset, and the bottom plots show the results for $1000$ (left) and $500$ (right) datasets. Here, red and blue lines correspond to the power spectrum prediction of the ANN and BNN emulators, respectively. The blue shaded region is the $3\sigma$ uncertainty associated with the mean prediction by BNN. The true power spectrum is shown with the black line. The bottom panel of each plot shows the percentage errors of the ANN prediction in red and the BNN mean prediction in blue.}
    \label{fig:PS_prediction}
\end{figure}

The performance of both of these emulators on an unseen test data for parameters $M_{(\rm h, min)} = 467.37 \times 10^8 \rm{M}_\odot$, $N_{\rm ion} = 181.05$, $R_{\rm mfp}=12.65\; \rm{Mpc}$ and mass average neutral fraction $\bar{x}_{\rm HI} = 0.377$ is shown in the figure \ref{fig:PS_prediction}. This particular choice of the parameter is to demonstrate the performance of our emulators at a low neutral fraction ($\bar{x}_{\rm HI} = 0.377$), which was absent in \cite{Tiwari_2022} due to a lack of training dataset. The true simulated values of the power spectrum, shown in black, are compared with the power spectrum predicted by the ANN-based emulator(in red) along with the mean prediction of the power spectrum by the BNN-based emulator (in blue). We also show the prediction uncertainty associated with the BNN-based emulator with $3\sigma$ confidence in the blue shaded region. We can see that the mean values of the power spectrum predicted by the BNN-based emulator are similar to or better than the point estimation for most of the $k$-bins provided by the ANN-based emulator. For all $k$-bins, the mean values predicted by the BNN-based emulator are within $3\sigma$ of the true power spectrum values, whereas there is no way to quantify the same for the power spectrum predicted by the ANN-based emulator.

Furthermore, as mentioned earlier, we perform the robustness study by systematically decreasing the total training dataset. The initial training was done using $7150$ training dataset. We randomly sample from this complete dataset to generate the smaller training dataset. Here, we show the results for the $1000$ and $500$ datasets, which we found as the lowest number of training datasets for which both ANN-based and BNN-based emulators were performing considerably well. Performance of emulators trained on these smaller datasets for the same unseen reionization parameters is shown in figure \ref{fig:PS_prediction}. We can see that as we decrease the training dataset, the point prediction of the ANN-based emulator deviates considerably from the true values of the power spectrum. This happens because for smaller training datasets, ANN-based emulators tend to overfit the training data and thus perform poorly on the unseen test data. Meanwhile, the BNN-based emulators perform better, as we can see that the uncertainties associated with the BNN predictions have increased; however, the true values are still within the $3\sigma$ confidence of the mean prediction. The BNN leverages Bayesian inference to estimate the network parameters. The epistemic uncertainty in BNN is mostly due to the uncertainty of the network parameters for the given training dataset. When we use a small training dataset, this epistemic uncertainty increases that resulting in an increase in overall prediction uncertainty.

\begin{table}[H]
\centering
\begin{minipage}[t]{0.47\linewidth}
\centering
\begin{tabular}{lcc}
    \toprule
    \textbf{Layer} & \textbf{Nodes} & \textbf{Activation} \\
    \midrule
    Input Layer & 3 & - \\
    Hidden Layer 1 & 1024 & ELU \\
    Hidden Layer 2 & 512 & ELU \\
    Hidden Layer 3 & 256 & ELU \\
    Hidden Layer 4 & 128 & ELU \\
    Hidden Layer 5 & 64 & ELU \\
    Hidden Layer 6 & 32 & ELU \\
    Output Layer & 7 & - \\
    \bottomrule
\end{tabular}
\vspace{0.5em}
\\ \small ANN-based emulator
\end{minipage}
\hfill
\begin{minipage}[t]{0.47\linewidth}
\centering
\begin{tabular}{lcc}
    \toprule
    \textbf{Layer} & \textbf{Nodes} & \textbf{Activation} \\
    \midrule
    Input Layer & 3 & - \\
    Hidden Layer 1 & 50 & ELU \\
    Hidden Layer 2 & 100 & ELU \\
    Hidden Layer 3 & 50 & ELU \\
    Output Layer & 7 & - \\
    \\
    \\
    \\
    \bottomrule
\end{tabular}
\vspace{0.5em}
\\ \small BNN-based emulator
\end{minipage}

\vspace{1em}
\caption{Model architecture of power spectrum emulators: ANN-based emulator (Left) and BNN-based emulator (Right)}
\label{tab:PS_ANN_BNN}
\end{table}

\subsection{Bispectrum emulation}
\begin{figure}[hbt]
    \centering
    \includegraphics[width=1\linewidth]{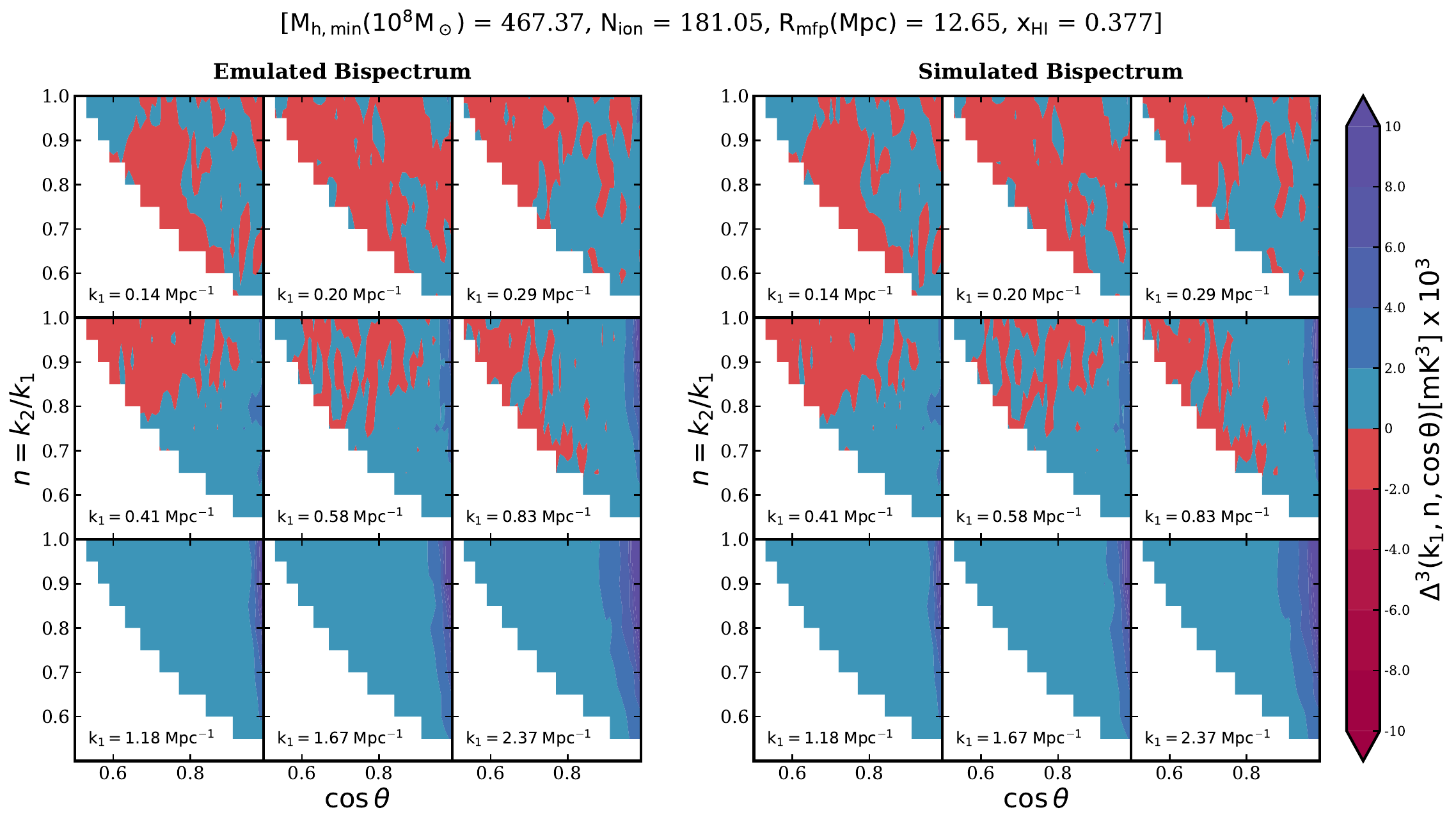}
    \includegraphics[width=0.5\linewidth]{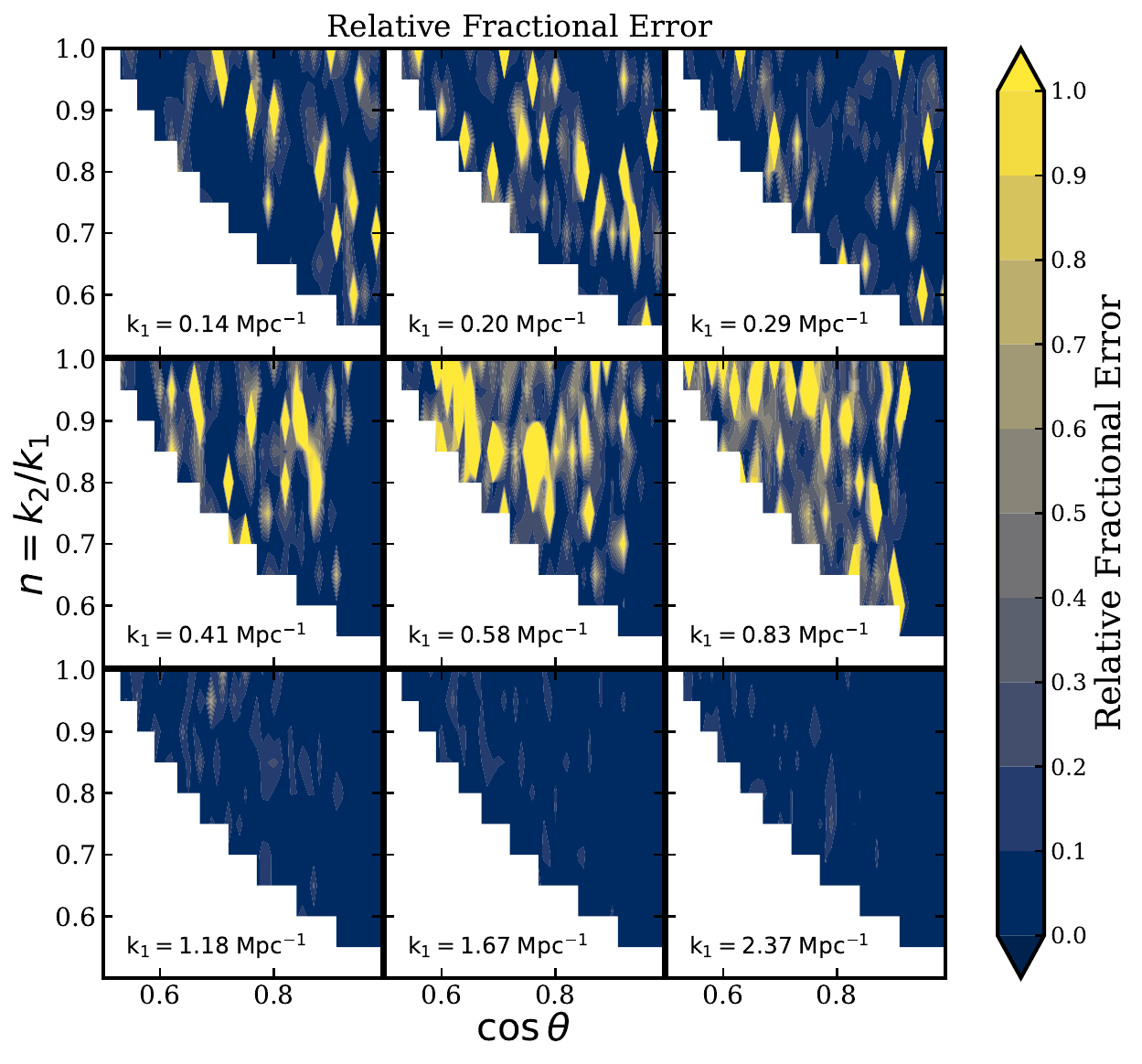}
    \caption{Bispectrum prediction by ANN emulators. The top left panel shows the emulated bispectrum, and the top right panel shows the simulated bispectrum for the same set of EoR parameters. The relative fractional error between the simulated and emulated bispectrum is shown in the lower panel. The islands of the high relative fractional errors mostly correspond to the errors in the sign prediction. }
    \label{fig:ANN_test1}
\end{figure}

We consider the bispectrum corresponding to a total of 9 $k_1$-bins in the range $[0.14-2.37]\; \rm{Mpc}^{-1}$. Similarly to the power spectrum, this range of $k_1$-bins is also chosen following the range of $k$-bins that upcoming telescopes will be able to probe. For each $k_1$-bin, we have $11$ bins for $n$ and $50$ for the $\cos{\theta}$ values. That gives us bispectrum values for a total of $550$ bins of triangle configurations. However, as mentioned earlier, if we consider only the unique triangle shapes ($ n\cos {\theta} \geq 1$), the number of triangle configurations reduces to $328$ for each $ k_1$-bin. Furthermore, unlike the power spectrum, which is always positive, the bispectrum can have both positive and negative values. Since we have a total of $9$ $k_1$-bins, we want to develop the emulator that predicts the sign and magnitude of a total of $(9 \times 328 =)\:2952$ triangle configurations of the bispectrum. The values of the bispectrum have a high dynamic range depending on the $k_1$-bins, and a single ANN architecture cannot capture the high dynamic range and sign variability of all $k_1$-bins for all parameters of the training dataset. So, to overcome this problem, we made nine different ANN emulators corresponding to $9$ $k_1$-bins, and each ANN emulator is trained to predict the magnitude and sign of the bispectrum via regression and classification, respectively. There are two main architectures for our ANN emulator. Both architectures have an input layer with three neurons and two output layers having $328$ neurons, one for sign and one for magnitude. We have used the ReLU activation function and mean squared error for the loss function during the training of all emulators. The learning rate is $\sim 10^{-3}$ for all emulators, but exact values differ for each emulator. The detailed architectures of these ANN emulators are provided in table \ref{tab:ANN_architectures}. The prediction of ANN trained on a full dataset compared to the simulated bispectrum is shown in figure \ref{fig:ANN_test1}.

\begin{table}[htbp]
\centering
\scalebox{0.95}{
\begin{minipage}{0.5\linewidth}
\centering
\begin{tabular}{lcc}
\toprule
\textbf{Layer} & \textbf{Nodes} & \textbf{Activation} \\ 
\midrule
Input Layer & 3 & - \\ 
Hidden Layer 1 & 2255 & ReLU \\ 
Hidden Layer 2 & 1524 & ReLU \\ 
Hidden Layer 3 & 1008 & ReLU \\ 
Output Layer 1 & & \\ (Magnitude) & 328 & - \\ 
Output Layer 2 & & \\ (Sign) & 328 & - \\ 
\\
\bottomrule
\end{tabular}
\vspace{0.5em}
\end{minipage}
\hfill
\begin{minipage}{0.5\linewidth}
\centering
\begin{tabular}{lcc}
\toprule
\textbf{Layer} & \textbf{Nodes} & \textbf{Activation} \\ 
\midrule
Input Layer & 3 & - \\ 
Hidden Layer 1 & 2255 & ReLU \\ 
Hidden Layer 2 & 1524 & ReLU \\ 
Hidden Layer 3 & 1008 & ReLU \\ 
Hidden Layer 4 & 956 & ReLU \\ 
Output Layer 1 & & \\ (Magnitude) & 328 & - \\ 
Output Layer 2 & & \\ (Sign) & 328 & - \\ 
\bottomrule
\end{tabular}
\vspace{0.5em}
\end{minipage}
}
\vspace{1em}
\caption{ANN model architecture of the two bispectrum emulators: left for the first four $k_1$-bins and right for the remaining five $k_1$-bins.}
\label{tab:ANN_architectures}
\end{table}

\begin{table}[htbp]
    \centering
    \begin{tabular}{lcc}
        \toprule
        \textbf{Layer} & \textbf{Nodes} & \textbf{Activation} \\
        \midrule
        Input Layer & 3 & - \\
        Hidden Layer 1 & 128 & ELU\\
        Hidden Layer 2 & 256 & ELU \\
        Hidden Layer 3 & 512 & ELU\\
        Output Layer & 328 & - \\
        \bottomrule
    \end{tabular}
    \caption{BNN model architecture of bispectrum emulator for all $k_1$-bins.}
    \label{tab:BNN_architecture}
\end{table}

\begin{figure}[htbp]
    \centering
    \includegraphics[width=1\linewidth]{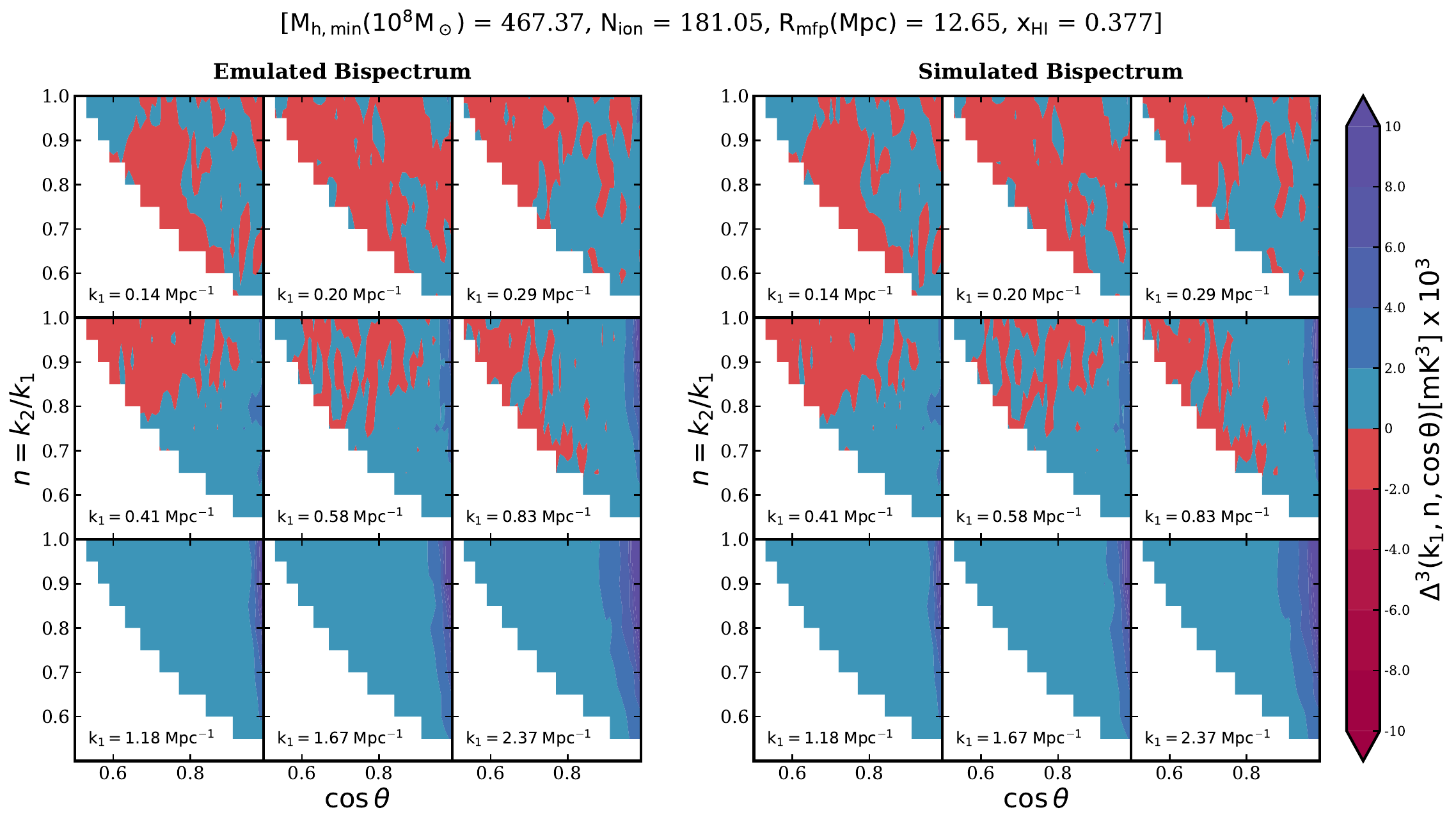}
    \includegraphics[width=0.5\linewidth]{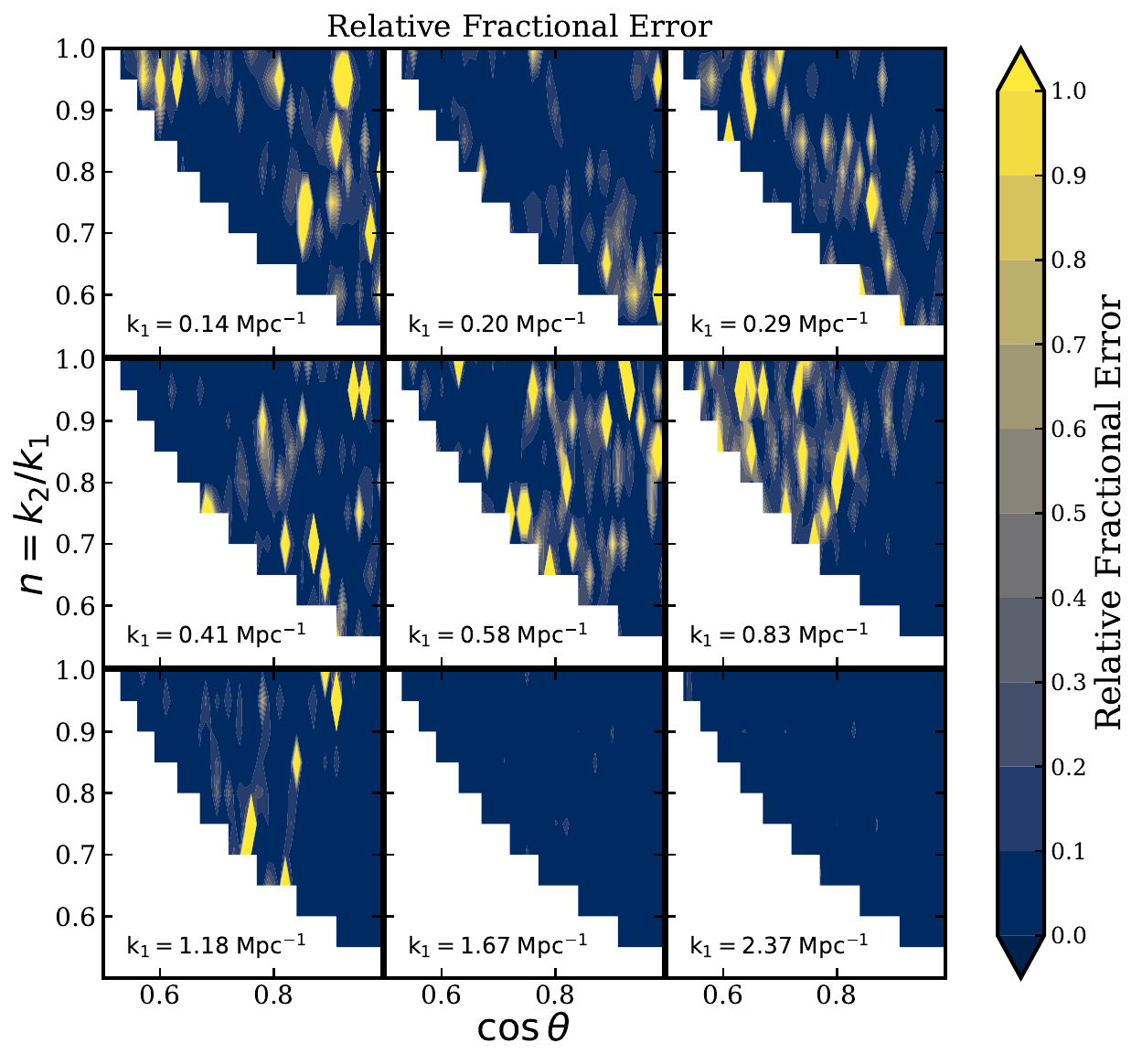}
    \caption{Bispectrum prediction by BNN emulators. The top left panel shows the mean values of the emulated bispectrum, and the top right panel shows the simulated bispectrum for the same set of EoR parameters. The relative fractional error between the simulated and emulated bispectrum is shown in the lower panel. We can see that the island of high relative errors is overall decreased compared to the ANN prediction.}
    \label{fig:BNN_test1}
\end{figure}

We train the BNN-based bispectrum emulator on the same datasets. Similar to the power spectrum, we require less depth in our BNN networks compared to ANN-based networks to capture more information about the bispectrum. Similar to ANN bispectrum emulators, we train $9$ BNN models for each $k_1$-bin. Each BNN has an input layer with three neurons and only one output layer with $328$ neurons; since we use Bayesian inference, we do not separate signs and magnitudes for BNNs. All BNNs have three hidden layers. We have used the ELU activation function for all the BNN architectures. We perform a total of $2000$ MCMC steps, with the initial $500$ steps being the warm-up steps to train the emulators. The performance of the BNN-based emulator is shown in figure \ref{fig:BNN_test1}. We compare the prediction of both the emulator using the relative fractional error. For both cases, there are islands of very high (some even more than $100 \%$) as evident from the figures \ref{fig:ANN_test1} and \ref{fig:BNN_test1}. These high relative fractional errors are mostly due to the wrong sign prediction by the emulators, as they occur for the triangles where there are clear sign variations in the emulated bispectrum. Also, the $k_1$-bins where there is no sign change, i.e., $k_1 = (1.18, 1.67, 2.37)~\rm Mpc^{-1}$, do not have these islands of high relative fractional error as bispectrum values for all the triangles are positive. BNN predicted bispectrum in general has fewer of these islands than the ANN prediction, which shows that training the neural network through Bayesian inference improves the performance of the network.

\begin{figure}[h]
    \centering
    \includegraphics[width=0.9\linewidth]{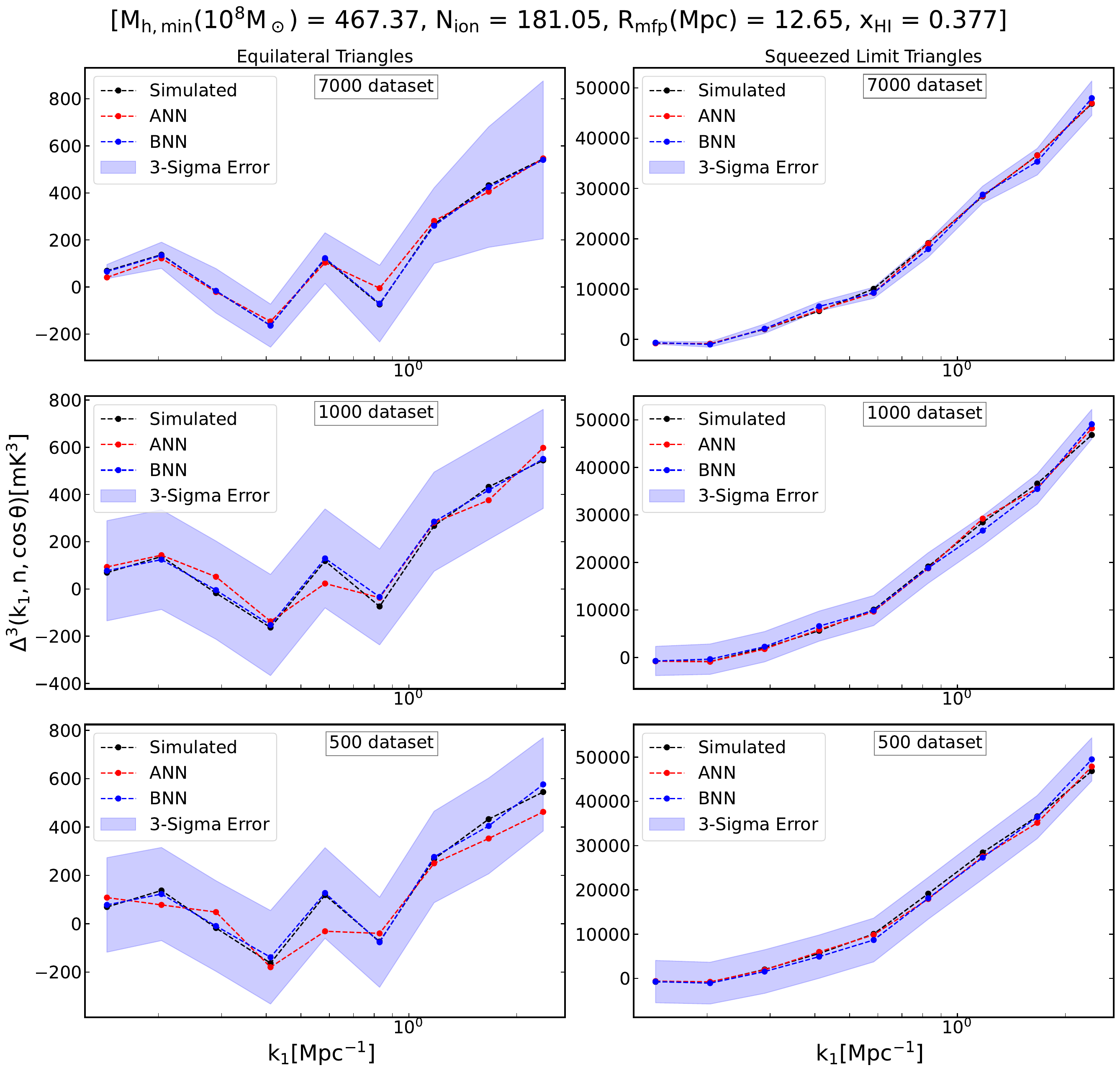}
    \caption{Line plots for the two triangle configurations of the bispectrum. The left column shows emulator predictions for equilateral triangle configurations, and the right column shows predictions for squeezed-limit triangle configurations. The red line shows the bispectrum prediction by the ANN, and the blue line shows the prediction by the BNN, with the blue shaded region representing the $3\sigma$ error on the BNN prediction. Both predictions are compared with the simulated bispectrum presented with the black line. The rows represent the size of training datasets used to train the corresponding emulators.}
    \label{fig:line-plots}
\end{figure}

Similar to the power spectrum, we checked the robustness of the BNN-based bispectrum emulators. For this, we trained our neural networks for a smaller dataset of $500$ and $1000$ randomly sampled parameter sets from the full dataset. The numbers of hidden layers are $[128, 256]$ and $[52, 128]$ for the $1000$ and $500$ datasets, respectively. The predictions by these networks are compared to the networks trained on the complete data, as shown in figure \ref{fig:line-plots}.
Here we have shown the predictions for two triangle configurations: equilateral and squeezed-limit triangles. The top row shows the prediction of the equilateral triangles (left) and squeezed-limit triangles (right) given by both ANN-based and BNN-based emulators. The shaded regions show the $3 \sigma$ errors associated with the prediction of the BNN-based emulators. The same is shown for emulators trained with $1000$ and $500$ datasets in the second and third rows, respectively. We can see that the $3\sigma$ errors associated with the BNN emulator prediction of the equilateral triangles are greater than the squeezed-limit triangle because the equilateral triangles have more sign flips for the bispectrum values, whereas the squeezed-limit bispectra remain positive always. Therefore, it is easier for emulators to train and predict the squeezed-limit bispectrum, and consequently, the errors associated with its prediction are less compared to the equilateral bispectrum. The BNN prediction errors also increase with a decrease in the training dataset for both triangle configurations. We can also see from the figure \ref{fig:line-plots} that the mean predictions of the BNN emulators (in blue) are better than the point predictions of the ANN emulators (in red), even though both are trained on the same dataset. This fact is also evident in the relative fractional error plot shown in figure \ref{fig:ANN_test1} and figure \ref{fig:BNN_test1}.

\section{Parameter Estimation}{\label{Results}}

Here, we constrain the reionization parameters using Bayesian inference and use ANN and BNN-based emulators for the power spectrum and bispectrum as forward models in the Bayesian inference pipeline to get the posterior of the parameters. Since we are interested in parameter estimation for a given reionization model, we can safely ignore model evidence as a normalizing constant. Then, posterior becomes proportional to the product of the likelihood $p(\mathcal{D} | \theta, \mathcal{M} ) $ and the prior $ p(\theta | \mathcal{M}) $, where $\theta$ is a parameter vector, $\mathcal{D}$ is observed data and $\mathcal{M}$ is model. We have assumed the uniform prior over all three parameters. A multivariate Gaussian log-likelihood is defined as:
\begin{equation}
    \rm{ln}~\mathcal{L} = \rm{ln}~p(\mathcal{D} | \theta, \mathcal{M}) = - \frac{1}{2} \big[ (P_{ref} - \mu)^T ~ (\Sigma)^{-1} ~ (P_{ref} - \mu) \big] - \frac{1}{2}~ ln ( 2\pi ~ |\Sigma|).
\end{equation}
Here $\rm P_{ref}$ denotes the observed data (for our case, it can be mock power spectrum or bispectrum data), $\mu$ denotes our model prediction (prediction of ANN or BNN emulators for respective statistic), and $\Sigma$ is the error covariance. 

\subsection{Estimation of Error Covariance}

Prior estimation of errors in the inference pipeline is necessary for robust likelihood estimation. The error covariance matrix has information about errors associated with the observation and, in the case of BNN-based emulators, errors associated with the model prediction. The observed 21-cm signal from EoR has uncertainties due to many observational effects, such as foreground contamination, sample variance, calibration errors, and instrumental noise. We consider only sample variance and instrumental noise for our analysis, assuming other contaminations are well-modeled and removed from the observed signal. Considering the 21-cm signal and system noise will not be correlated with each other, we can write the total covariance as: 
\begin{equation}
    \Sigma = \sigma_{\rm T}^2  = \sigma^2_{\rm SV} + \sigma^2_{\rm N},
\end{equation}
$\sigma^2_{\rm SV}$ and $\sigma^2_{\rm N}$ are covariance contributions from sample variance and system noise, respectively. These covariance calculations are different for the power spectrum and bispectrum. For the power spectrum, we calculate the sample variance from the binned power spectrum following:
\begin{equation}
    \sigma^2_{\rm SV}(P_i) = \frac{\Big(\Delta_b^2(k_i)\Big)^2}{N_{k_i}},
\end{equation}
where $\Delta_b^2(k_i)$ and $N_{k_i}$ are the power and total sample of $k$-bins in the $i^{\rm th}$ bin, respectively. Similar to the power spectrum, the sample variance of the bispectrum is obtained from the bin-averaged bispectrum $  \bar{B}_i$$(k_1,n,\cos{\theta})$ from,
\begin{equation}
     \sigma^2_{\rm SV} (B) = \frac{\Big(\Delta_b^3(k_1,n,\cos{\theta})\Big)^2}{N_{\rm tri}},
\end{equation}
here $\Delta_b^3(k_1,n,\cos{\theta})$ is normalized binned bispectrum, and $N_{\rm tri}$ is the total number of triangles in $i^{th}$ triangle bin. The instrumental noise contribution for the power spectrum can be calculated as follows:
\begin{equation}
    \sigma^2_{\rm N} (P_i) = 2\;\frac{V_{f}}{V_{P}}\; \Big(\Delta_{\rm N}^2(k_i)\Big)^2.
\end{equation}
Here $V_f = 2\pi^3/V_s$ is the volume of the fundamental cell in the Fourier domain, where $V_s$ is the total survey volume,  $V_P = 4\pi k_i\;\Delta k_i$ is the shell volume of the $i^{th}$ bin with $\Delta k_i$ bin width and $\Delta_{\rm N}^2(k_i) = k_i^3 P_{\rm N}(k_i) / (2\pi^2)$ is normalized noise power in the $i^{th}$ bin. The noise power spectrum $P_{\rm N}(k_i)$ for a radio interferometer is given as \cite{Bull_2015, Obuljen_2018}:

\begin{equation}
    P_{\rm N}(k,z) = \frac{T^2_{\rm sys}(z) \chi^2(z) r_{\nu}(z) \lambda^4(z) }{ A^2_{\rm eff} t_{\rm obs} N_{\rm pol} \Sigma(\bm{U},z) \nu_{\rm 21cm} }.
\end{equation}
Here, $\lambda(z) = 21 \times (1+z) ~\rm cm$ and $\nu_{\rm 21cm} = 1420\;\rm{MHz}$ are the wavelengths at redshift $z$ and restframe wavelength of the 21-cm signal. The $\chi(z)$ is  comoving distance to redshift $z$, and $r_{\nu} = (c/H(z))(1+z)^2$. The system temperature $T_{\rm sys}$ can be calculated using $T_{\rm sys}(\nu) = 100 + 300(150 \; \rm{MHz} / \nu)^{2.55} \;K$ \cite{Mellema_2013}. We consider the number of polarizations ($N_{\rm pol}$) to be 2. The effective collecting area of each antenna is given as $A_{\rm eff} = A_{\rm eff} (\nu_{\rm crit}) \times \epsilon(\nu)$, where $\epsilon (\nu)$ is defined as \cite{Bull_2015}
\begin{equation}
    \epsilon(\nu) =
     \begin{cases}  
     (\nu_{\rm crit}/\nu)^2, & \text{if } \nu > \nu_{\rm crit}\\
    1, & \text{otherwise}
     \end{cases}
\end{equation}
The baseline density of system $\Sigma(\bm{U},z) = N_a^2 / (2\pi U_{\rm max}^2)$ is considered to be constant within the core radius. $N_a$ is the total number of such antenna and $u_{\rm max}$ is the maximum baseline in the units of $\lambda(z)$. We can estimate the error covariance for the bispectrum using different methods \cite{Yoshiura_2015, Mondal_2021, Murmu_2024}. We follow \cite{Murmu_2024} for the variance calculation of the bispectrum. The variance in the bispectrum due to system noise is computed using \cite{Scoccimarro_2004, Liguori_2010},
\begin{equation}
    \sigma_{\rm N}^2(B) \approx s_B ~ \frac{V_f}{V_B} P_{\rm N}(k_1,z)P_{\rm N}(k_2,z)P_{\rm N}(k_3,z).
\end{equation}
Here, $V_f$ and $P_{\rm N}(k,z)$ are the same volumes of the fundamental cell and noise power spectrum as mentioned above for the power spectrum. The values of $s_B$ are 1, 2 and 6 for general, isosceles and equilateral $k-$bins triangles, respectively, and $V_B \approx 8\pi^2k_1k_2k_3\Delta k_1 \Delta k_2 \Delta k_3$, where $k_1, k_2$ and $k_3$ are the length of the sides of the bispectrum triangle in the $k$-space bin, and the bin width of those $k$ values are $\Delta k_1, \Delta k_2,\Delta k_3$, respectively. For this work, we have considered $1000$ hr of SKA-LOW observation and all the parameters values related to instumental noise covariance for both power spectrum and bispectrum are taken as follows: $A_{\rm eff}(\nu_{\rm crit})$ is taken to be $962\;\rm{m^2}$ with $\nu_{\rm crit}$ being $\rm{110 \; MHz}$ \cite{Giri_2018}. We consider core radius of $R_{\rm max} = 2\;\rm km$ with total antenna $N_a = 296$ \cite{Mazumdar_2022}. For the calculation of $V_B$ for the bispectrum, the relative $k$-bin size is taken as $\Delta k/k \sim 1$.

\begin{figure}[ht]
    \centering
    \includegraphics[width=0.9\linewidth]{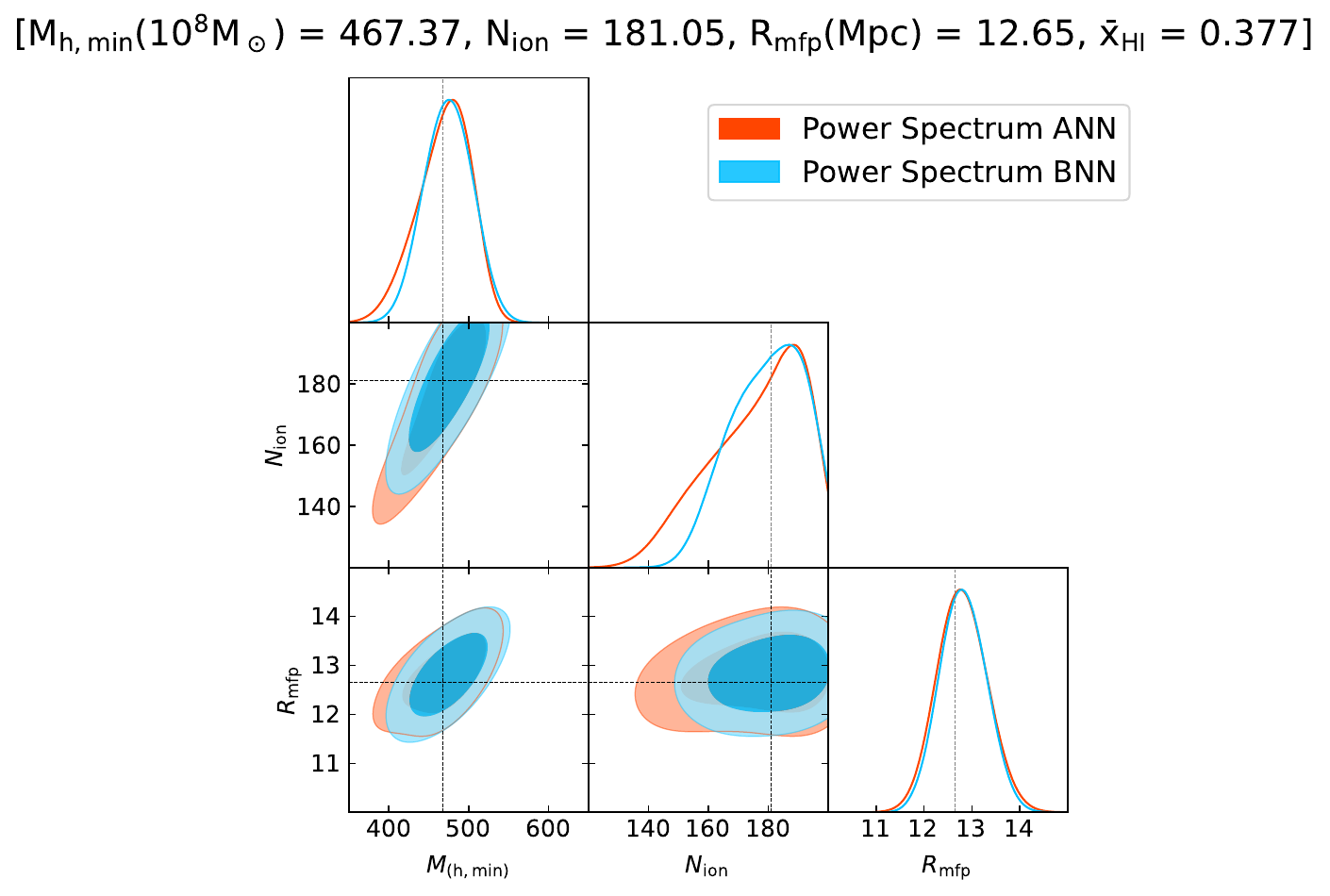}
    \includegraphics[width=0.48\linewidth]{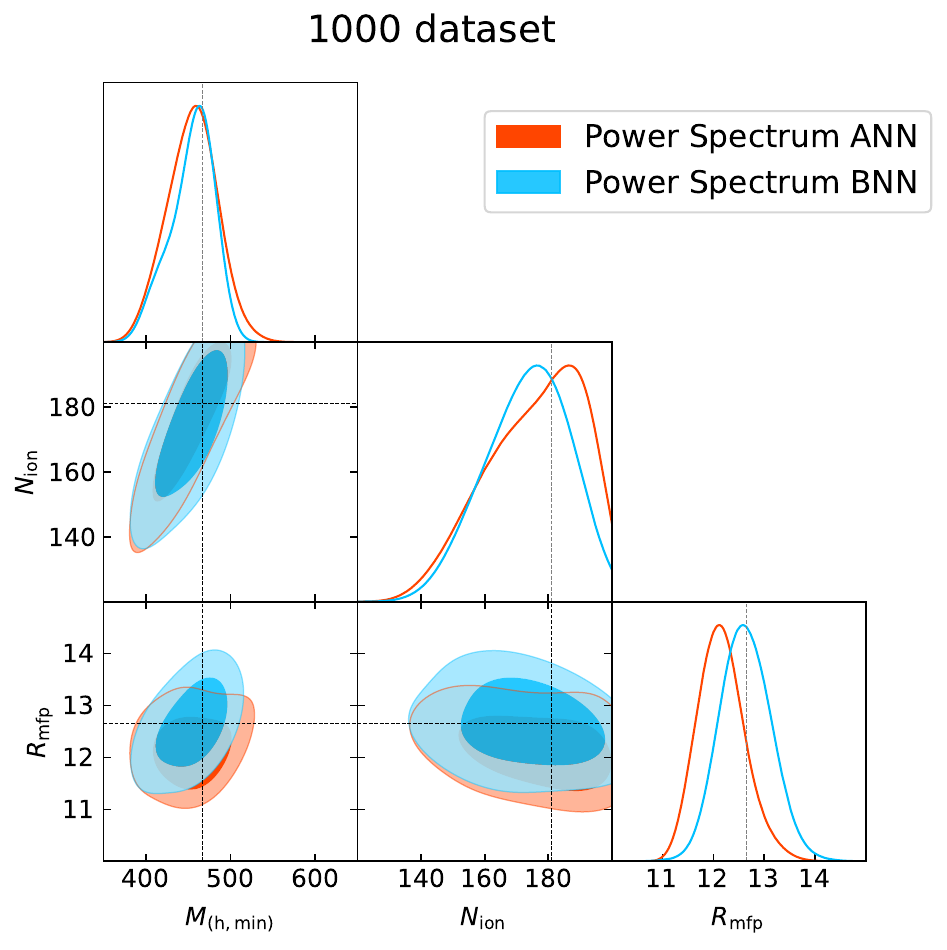}
    \includegraphics[width=0.48\linewidth]{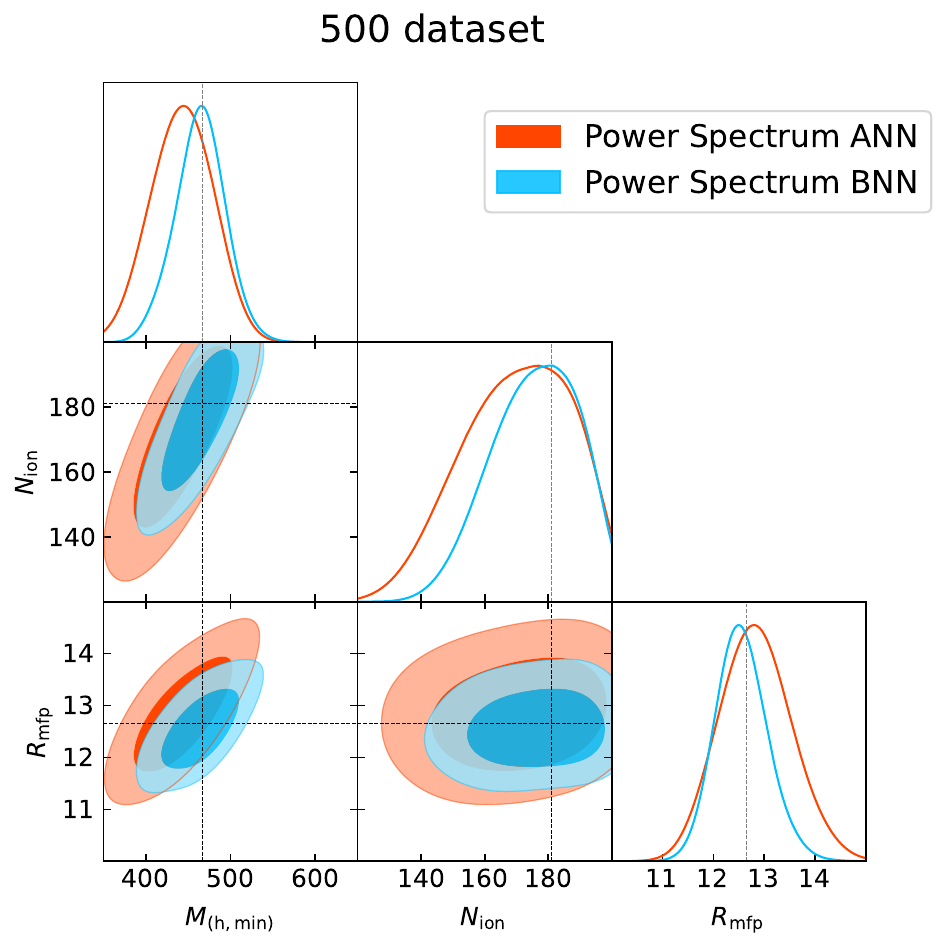}
    \caption{Posteriors of reionization parameters obtained using power spectrum as a summary statistic, emulated using ANN-based emulator (orange) and BNN-based emulator (sky blue). The top panel shows the constraints from emulators trained on the complete $(7000)$ dataset. In the bottom panel, the left plot shows constraints from emulators trained on 1000 data points, and the right one from 500 data points. The true reionization parameters $M_{(\rm h, min)} = 467.37 \times 10^8 \rm{M}_\odot$, $N_{\rm ion} = 181.05$, $R_{\rm mfp}=12.65\; \rm{Mpc}$ and $\bar{x}_{\rm HI} = 0.377$ are shown using black dotted lines. }
    \label{fig:PS_ANNvsBNN}
\end{figure}

\subsection{Results}
Here, we present and compare the estimation of the reionization parameters provided by both types of emulators. We have shown the corner plots that compare the posterior of parameters estimated via ANN and BNN emulators. This comparison is done using emulators trained on three different sizes of datasets (as discussed in Sec \ref{Emulators}) for both the power spectrum and the bispectrum. Furthermore, we also compare the estimations obtained by the power spectrum and the bispectrum using the BNN-based emulator. The inference takes $\sim3.5$ hr for ANN and $\sim6$ hr on for $100$k samples using the NVIDIA A100-SXM4-40GB GPU. 

The reionization parameter estimation using power spectrum as a summary statistic is shown in figure \ref{fig:PS_ANNvsBNN}. The top panel shows the corner plot of the parameter estimation obtained via emulators trained on the complete datasets, whereas the two plots in the bottom panel show the estimation obtained using emulators trained on $1000$ datasets (left) and $500$ datasets (right). The diagonal panels in each of the plots show the marginalized distribution of the parameters given using the ANN-based emulator (in orange) and the BNN-based emulator (in skyblue), with true values represented using the dashed lines. The non-diagonal panels show the $2\text{D}$ joint distribution of the two parameters, where $1\sigma$ and $2\sigma$ contours are represented by dark and light-shaded regions, respectively, and true values are shown using vertical and horizontal lines. We can see that for all three cases, the BNN-based emulator provides better constraints than the ANN. For smaller datasets, ANN-based emulators provide biased posteriors than BNN. This happens because ANNs are more prone to overfitting and provide biased predictions for smaller training datasets. On the other hand, BNNs do not give biased predictions but predict higher uncertainties when trained on a smaller training dataset, as shown in figure \ref{fig:line-plots}. 

\begin{figure}[htbp]
    \centering
    \includegraphics[width=0.9\linewidth]{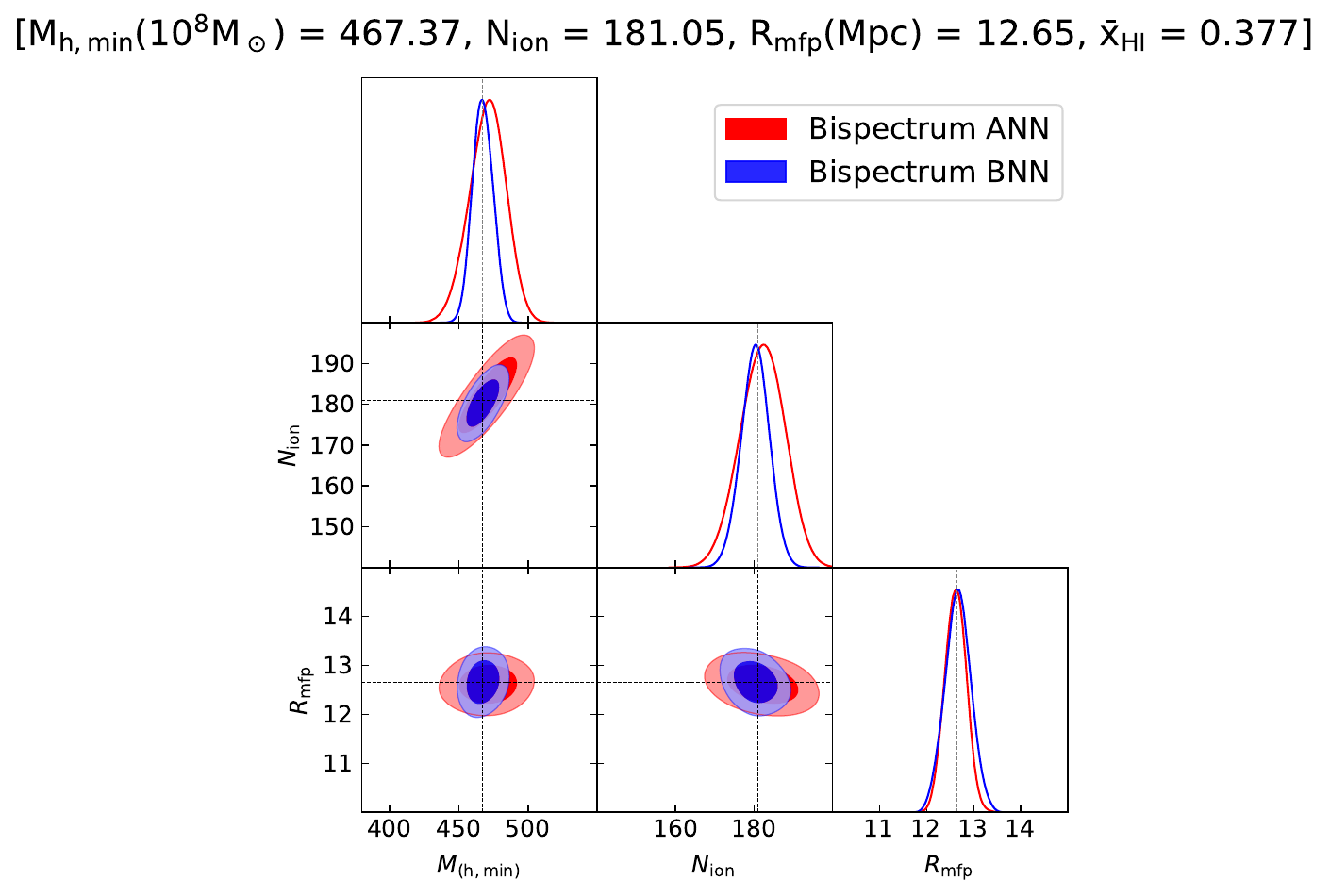}
    \includegraphics[width=0.48\linewidth]{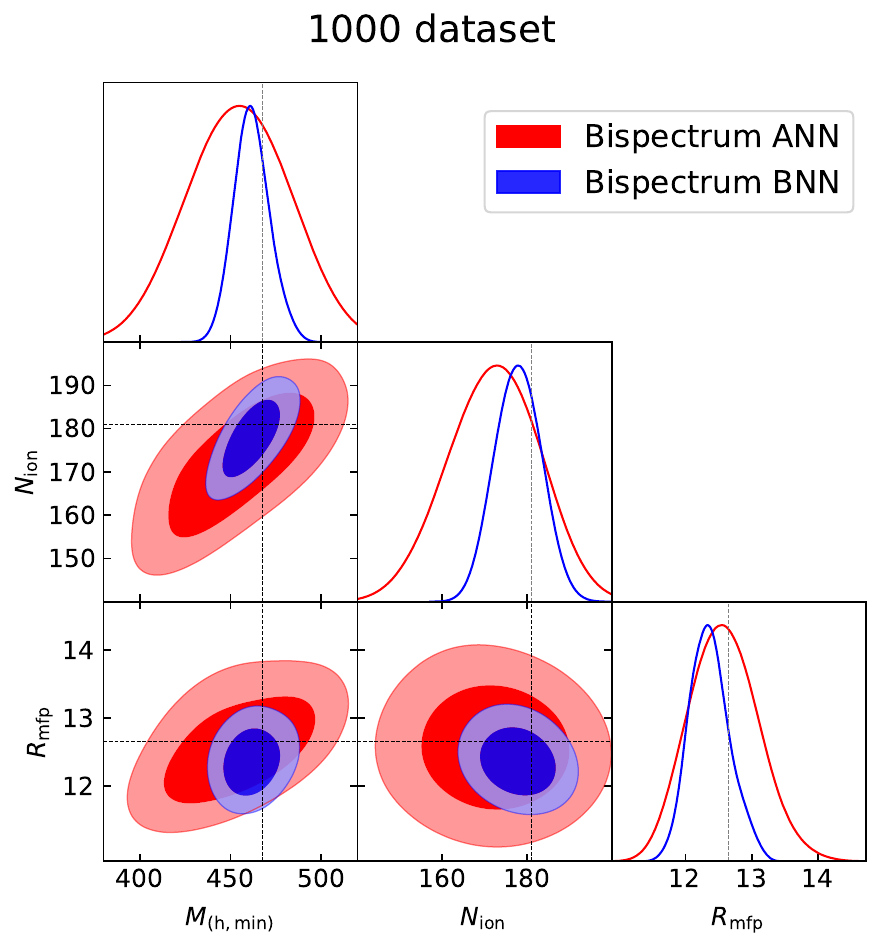}
    \includegraphics[width=0.48\linewidth]{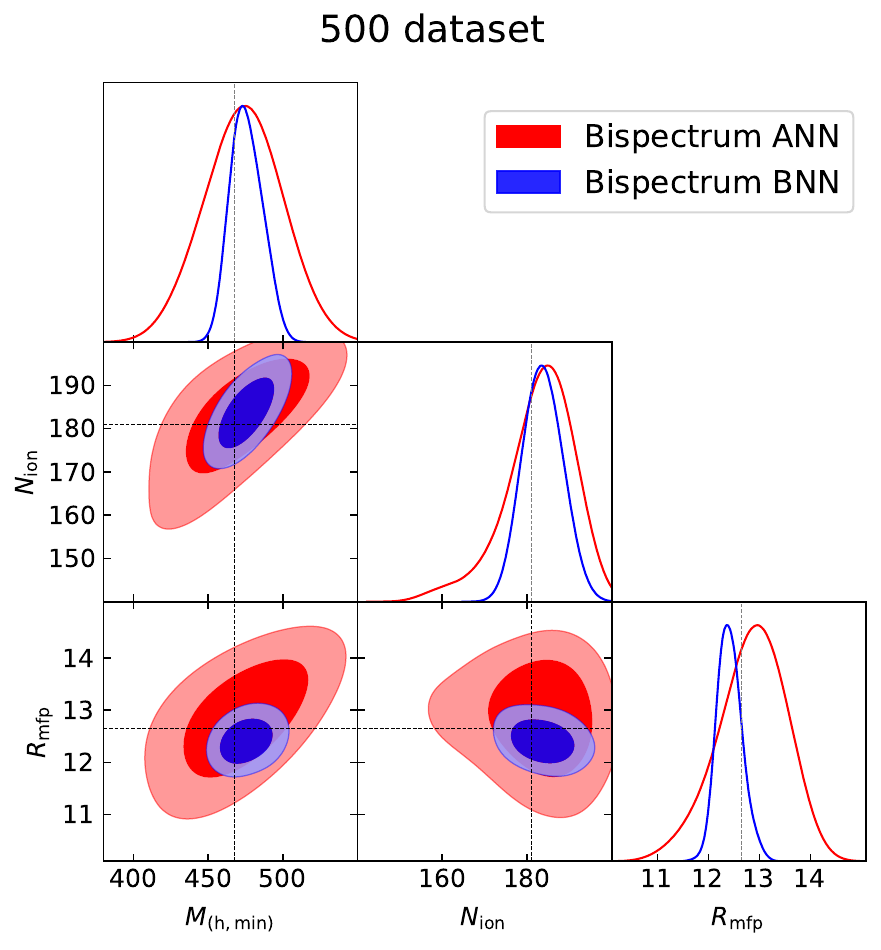}
    
    \caption{Posteriors of reionization parameters obtained using bispectrum as a summary statistic, emulated using ANN-based emulator (red) and BNN-based emulator (blue). The top panel shows the constraints from emulators trained on the complete dataset. In the bottom panel, the left plot shows constraints from emulators trained on 1000 data points, and the right one from 500 data points. The true reionization parameters $M_{(\rm h, min)} = 467.37 \times 10^8 \rm{M}_\odot$, $N_{\rm ion} = 181.05$, $R_{\rm mfp}=12.65\; \rm{Mpc}$ and $\bar{x}_{\rm HI} = 0.377$ are shown using black dotted lines.}
    \label{fig:BS_ANNvsBNN}
\end{figure}

Similar corner plots of the reionization parameter posteriors obtained using bispectrum as a summary statistic are shown in figure \ref{fig:BS_ANNvsBNN}. We have used all unique triangle configurations to estimate the reionization parameters. As in the power spectrum, here also, there are three plots showing the estimation obtained via emulators trained on three training datasets. We see that BNN-based emulators provide a tighter estimation of parameters than ANN-based emulators. This is because the bispectrum is a complex statistic to emulate as compared to the power spectrum. Bispectrum has a high dynamic range and sign variation, which are difficult for ANN-based emulators to emulate, especially for smaller datasets. BNN-based emulators can predict these variations in sign and magnitude at the cost of assigning a large error in their predictions for a given triangle configuration, as shown in figure \ref{fig:line-plots}. Therefore, for large datasets, ANN emulators do a good job as their constraints are comparable to those provided by BNN emulators; however, the difference in their estimation is very prominent for smaller datasets. So, BNN-based emulators overall perform better in emulating the bispectrum than ANN-based emulators and thus, provide better constraints of reionization parameters. 
Furthermore, it is important to notice that BNN-based emulators provide the distribution of the summary statistics given the input parameters, whereas ANN-based emulators provide only a single value of statistics sampled from that distribution. So, the BNN-based emulator inherently carries more information about the emulated statistics than the ANN-based emulator. Moreover, as we see in the in Sec \ref{Simulation} figure \ref{fig:line-plots}, even the mean values predicted by the BNN-based emulators are more accurate than the point prediction of the ANN-based emulators, especially for the case of smaller training datasets and for the triangle configurations of bispectrum which varies in sign for different $k_1$-bins. These result in a tighter constraint on reionization parameters using BNN-based emulators compared to ANN-based emulators for both power spectrum and bispectrum.\\

\begin{figure}[htbp]
    \centering
    \includegraphics[width=0.9\linewidth]{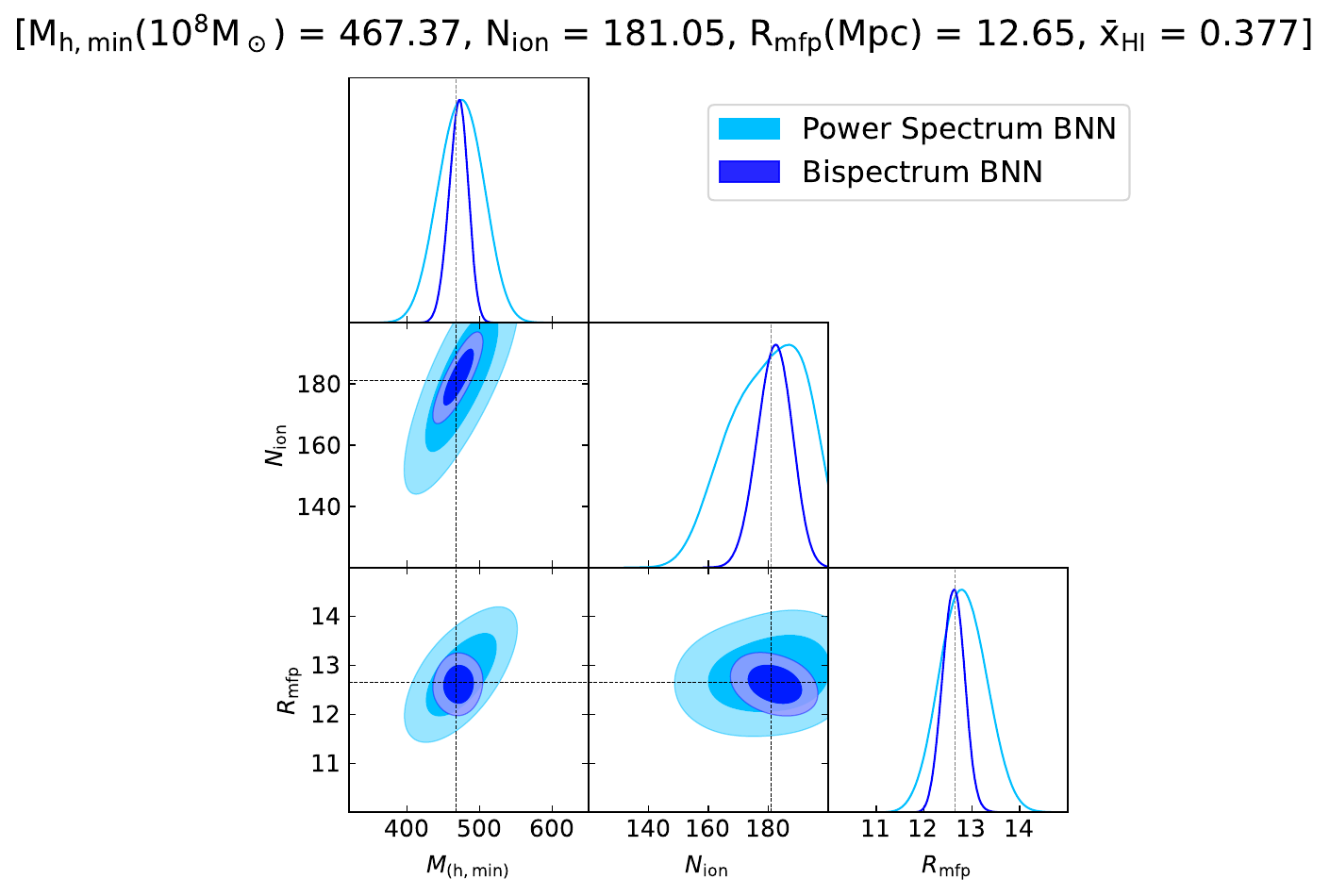}
    \includegraphics[width=0.48\linewidth]{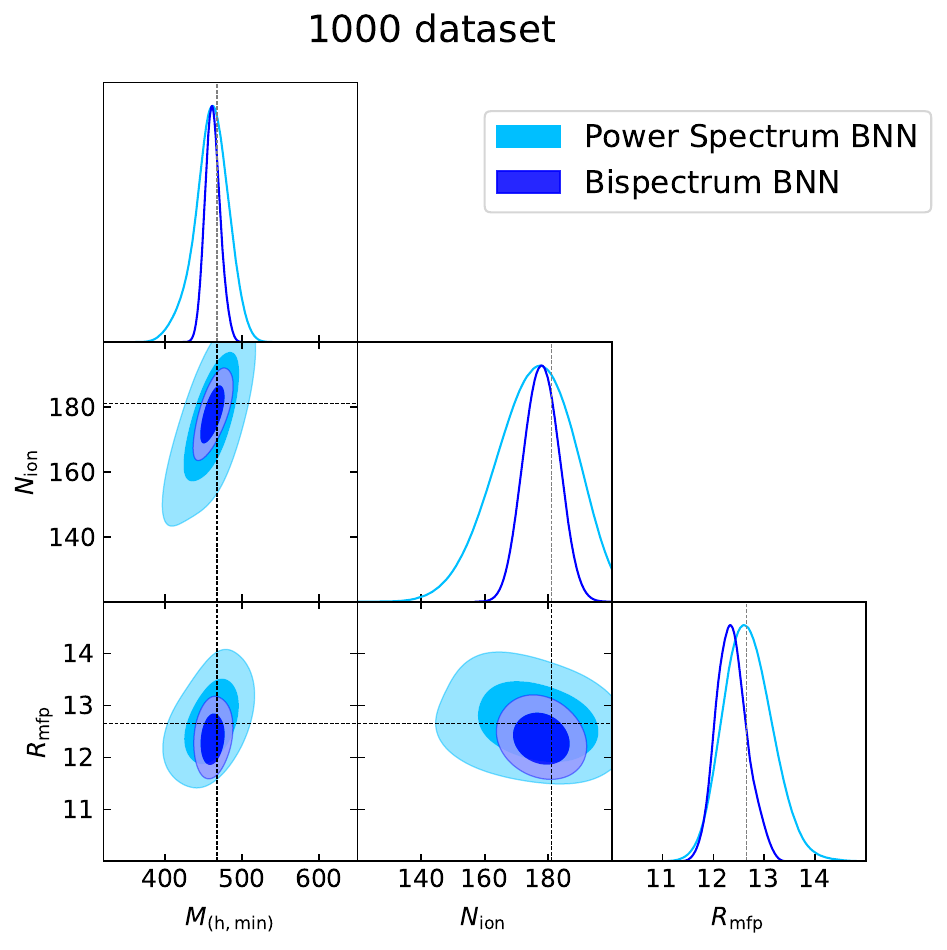}
    \includegraphics[width=0.48\linewidth]{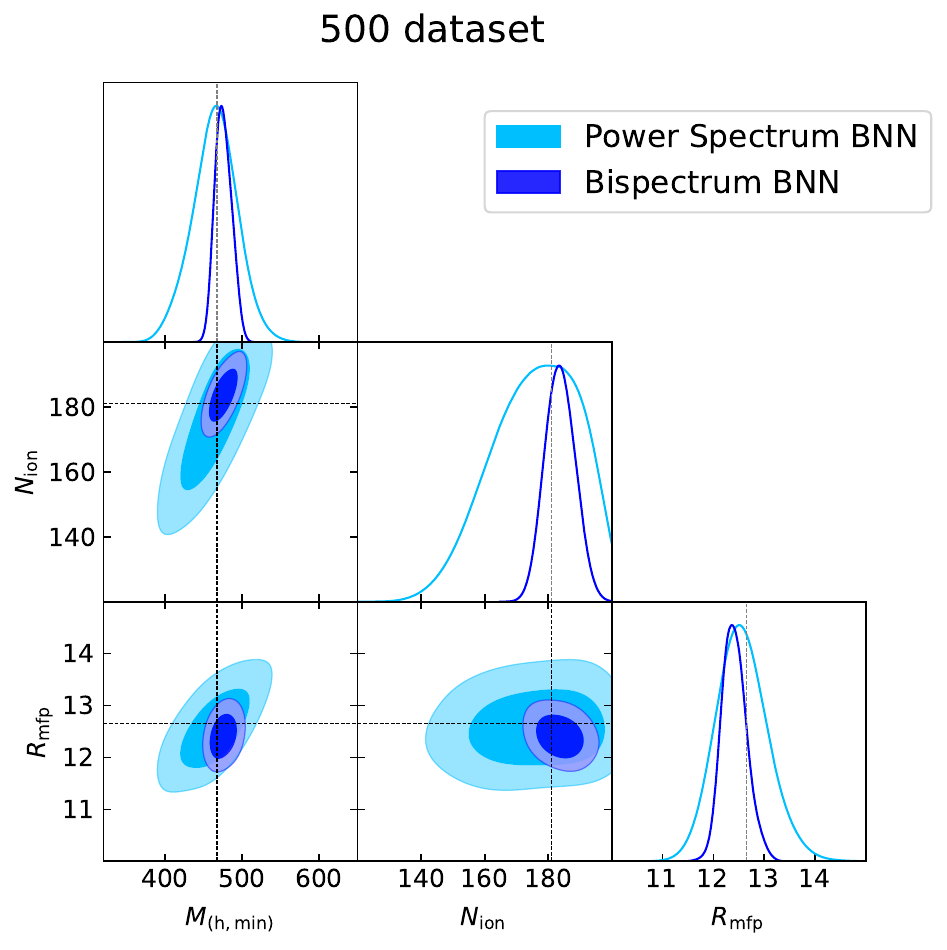}
    \caption{Comparison of constraining power of the power spectrum (sky blue) and bispectrum (blue) as a summary statistic emulated using BNN-based emulators for the true reionization parameters $M_{(\rm h, min)} = 467.37 \times 10^8 \rm{M}_\odot$, $N_{\rm ion} = 181.05$, $R_{\rm mfp}=12.65\; \rm{Mpc}$  and $\bar{x}_{\rm HI} = 0.377$ shown using black dotted lines.}
    \label{fig:PSvsBS}
\end{figure}

Additionally, we also compare the parameter constraining capabilities of two statistics, power spectrum and bispectrum, using the BNN-based emulator demonstrated in figure \ref{fig:PSvsBS}. It is evident from these plots that we get much tighter constraints on reionization parameters using the bispectrum as a summary statistic for all three cases. This is expected because, as discussed earlier, the bispectrum, being a higher-order (three-point) statistic, has more information content and consequently more constraining power than the power spectrum. Thus, considering the bispectrum for all unique triangle configurations ensures better constraints on the reionization parameters. These results are consistent with the findings of \cite{Tiwari_2022}, who compared the constraining power of the power spectrum and bispectrum of the EoR 21-cm signal using ANN-based emulators alone.

\section{Summary and Discussion}{\label{Summary}}

Many ongoing and upcoming radio interferometers, such as uGMRT, HERA, LOFAR, MWA, and SKAO, are trying to probe the EoR through the redshifted 21-cm signal with unprecedented accuracy and resolution. This 21-cm signal is a highly non-Gaussian signal; thus, probing this non-Gaussianity will provide a better understanding of the EoR physics. Therefore, the bispectrum is an optimal statistic to study and constrain the EoR parameters than the power spectrum. We use a Bayesian inference pipeline to obtain the statistical inference of the reionization model parameters from these summary statistics. This inference technique requires modeling the 21-cm signal using numerical or seminumerical simulations, and rerunning these simulations a large number of times is computationally expensive. To overcome this problem, neural network-based emulators are used to emulate the 21-cm signal statistics given reionization parameters.

Previous works on reionization parameter estimation using Artificial neural network (ANN) emulators for 21-cm signal summary statistics has a fundamental drawback that these ANN-based emulators can produce only point value predictions of the target signal statistic; thus, they fail to capture the uncertainty related to their own predictions. Therefore, when such emulators are used in the Bayesian inference pipeline, they can not naturally propagate their prediction uncertainties to the estimated model parameters. To address this problem, we have developed Bayesian neural network (BNN) based emulators for the power spectrum and bispectrum of the redshifted 21-cm signal. The BNN-based emulators provide the mean prediction for the summary statistics as well as the uncertainties associated with their prediction, and we can propagate this uncertainty through our Bayesian inference pipeline. We have demonstrated that using BNN-based emulators, we get a more robust and trustworthy estimation of the reionization parameters compared to the ANN-based emulators.

In this work, we show that the BNN-based emulator performs better than the ANN-based emulator, especially for the bispectrum, as it is a more complex statistic to emulate. Moreover, we systematically reduce the size of the training dataset to check the robustness of the emulators on the smaller datasets. The performance of the emulators trained with $1000$ and $500$ datasets is shown in Sec \ref{Emulators}. The prediction of the ANN-based emulators deviates considerably from the true values of the statistics, as it tends to overfit for the smaller dataset, whereas the BNN-based emulators provide better mean value predictions than the ANN. Also, the uncertainties associated with the BNN prediction increase with decreasing training dataset, and the true value of the statistics almost always lies within the $3\sigma$ of the mean prediction, whereas there is no quantification of the same for ANN emulators.

We use these emulators in a Bayesian inference pipeline to obtain the constraints on the reionization parameters. We show that BNN-based emulators provide more robust constraints for both power spectrum (see figure~\ref{fig:PS_ANNvsBNN}) and bispectrum (see figure~\ref{fig:BS_ANNvsBNN}) for all three training datasets. BNN-based emulators provide the distribution of summary statistics based on input parameters, whereas ANN-based emulators provide simply a single value of statistics drawn from that distribution. Therefore, compared to ANN-based emulators, BNN-based emulators offer more information about the emulated statistics. Furthermore, as shown in the figures in Sec \ref{Simulation}, even the mean values predicted by BNN-based emulators outperform the point prediction by ANN-based emulators, particularly when dealing with smaller training datasets and more sign flips for the bispectrum values. As a result, for both power spectrum and bispectrum, BNN-based emulators have better and more robust constraints on reionization parameters than ANN-based emulators. Moreover, we also show that we get better constraints on reionization parameters, considering the bispectrum (all unique triangle configurations) as a summary statistic, than the power spectrum for BNN-based emulators. This is in agreement with the results of \cite{Tiwari_2022}, where the same results were obtained through ANN-based emulators. Results of both of these works are different from the constraints obtained from some previous article (e.g. \cite{Watkinson_2022}). As explained in \cite{Tiwari_2022}, the bispectrum used in \cite{Watkinson_2022} is normalized with the power spectrum, whereas we do not use such normalization of bispectrum values. This allows us to exploit the large dynamic range of the bispectrum amplitude in this work. Furthermore, \cite{Watkinson_2022} only considers the bispectrum for the isosceles triangles, whereas here we consider all unique shapes of $\bm{k}$-triangles, thus maximising the information content in this statistic. In \cite{Tiwari_2022}, it is shown that considering a single $\bm{k}$-triangle configuration (specifically, isosceles) provides poorer constraints than the power spectrum; however, when one considers all the unique $\bm{k}$-triangle configurations, one obtains better constraints.

For this work, we have used coeval cubes at redshift of $\bm{z=7}$ instead of lightcone cubes due to the limitation of computational resources to generate the training dataset for the lightcones. We will explore the effect of the redshift evolution of the bispectrum to constrain the reionization parameters in future work. Furthermore, we developed the emulators for a specific inside-out reionization model used in \texttt{ReionYuga}. One of the drawbacks of the emulator-based inference pipeline is that emulators trained on a specific reionization model will make biased inference when applied to reionization models based on different astrophysical assumptions. Therefore, changing the reinoization model requires retraining the emulators on the training data generated using that particular model. The computational cost of the training will depend on the number of parameters of the model.

In this work, we assume that the signal is obtained after removing the effects of foreground contamination and does not contain any residual foreground. However, we have included the sample variance and telescopic noise for the upcoming SKA-LOW telescope for $\bm{1000}$ hr of observation for both the power spectrum and the bispectrum in our study, and assume that other systematics effects are removed from the signal. This is a simplified assumption, so a more realistic covariance would be larger than what we have assumed in this work. In future works, we plan to include the effects of lightcone, residual foreground, along with cosmic variance, and study their effects on the estimation of the reionization parameters using a BNN-based bispectrum emulator.

Recently, there have been many studies using simulation-based inference (SBI) to estimate the reionization parameters. One way to utilize the SBI is to estimate the reionization parameters directly from the maps using different neural architectures \cite{Gillet_2019, Hortúa_2020, Hassan_2020, Zhao_2022a, Prelogovic_2022, Neutsch_2022, Schosser_2025, Ore_2025}; however, these neural networks also compress the signal into latent summary space, and this comes with the additional challenge of physical interpretation of these compressed summaries. Alternatively, one can use the summary statistics to estimate the reionization parameters using SBI \cite{Zhao_2022b, Prelogovic_2023, Meriot_2025, Greig_2024}. Most of these studies with SBI use the power spectrum or other compressed statistics to estimate the reionization parameters, bypassing the need for the likelihood of such statistics. Since the 21-cm signal is a non-Gaussian signal, it is crucial to understand the informativeness of the different statistics in constraining the reionization parameters \cite{Prelogovic_2024, Sui_2023}. In the future, we would like to perform a similar analysis using the bispectrum to investigate the effect of higher-order statistics in estimating the reionization parameters using SBI.

\section*{Acknowledgments}

YM acknowledges the financial support by the Department of Science and Technology, Government of India, through the INSPIRE Fellowship. SM would like to thank the Science and Engineering Research Board (SERB) and the Department of Science and Technology (DST), Government of India, for financial support through Core Research Grant No. CRG/2021/004025 titled “Observing the Cosmic Dawn in Multicolour using Next Generation Telescopes”. LN acknowledges the financial support by the Department of Science and Technology, Government of India, through the INSPIRE Fellowship. CSM would like to acknowledge financial support from the Council of Scientific and Industrial Research (CSIR) via a CSIR-SRF Fellowship (Grant No. 09/1022(0080)/2019-EMR-I) and from the ARCO Prize Fellowship. AKS is supported by the National Science Foundation (grant no. 2206602). SD acknowledges the Cambridge Trust and Isaac Newton Studentship for funding his PhD.


\bibliographystyle{JHEP}
\bibliography{biblio.bib}


\end{document}